\documentclass[preprint,3p,times]{elsarticle}

\usepackage{amssymb}
\usepackage{amsmath,cases,extarrows}
\usepackage{mathrsfs}
\usepackage{graphicx}
\usepackage{amsthm}
\usepackage{bm}
\usepackage{enumitem}
\usepackage{algorithm}
\usepackage{algpseudocode}
\usepackage{hyperref}
\hypersetup{hypertex=true,
colorlinks=true,
linkcolor=blue,
anchorcolor=blue,
citecolor=blue}
\usepackage[nameinlink]{cleveref}

\usepackage{bm}
\usepackage{mdframed}
\usepackage{listings}
\usepackage{xcolor}

\newtheorem{theorem}{Theorem}[section]
\newtheorem{proposition}[theorem]{\indent Prop}
\newtheorem{lemma}[theorem]{\indent Lemma}
\newtheorem{corollary}[theorem]{\indent Cor}

 \newdefinition{remark}{Remark}
\newcommand{\ii}{\mathrm{i}}

\newtheorem{definition}{\indent Def}[section]

\newtheorem{assumption}{Assumption}
\numberwithin{equation}{section}

\journal{Journal of Computational Physics}

\begin{document}

\begin{frontmatter}



\title{Mathematical Analysis of Symmetry-Protected Bound States in the Continuum in Waveguide Arrays}


\author[1]{Xin Feng} 
\ead{fengxin24@mails.jlu.edu.cn}
\author[2]{Wei Wu\corref{cor1}}
\ead{wei\_wu@jlu.edu.cn}
\affiliation[1,2]{organization={School of Mathematics, Jilin University},
            addressline={Qianjin Street 2699},
            city={Changchun},
            postcode={130012},
            state={Jilin},
            country={China}}
\cortext[cor1]{Corresponding author}
\begin{abstract}
This paper presents a rigorous mathematical analysis for symmetry-based Bound States in the Continuum (BICs) in optical waveguide arrays. Different from existing research, we consider a finite system of horizontally and equidistantly aligned waveguides and transform the wave propagation problem into Nonorthogonal Coupled-Mode Equations (NCME), rather than adopting the tight-binding approximation or orthogonal coupled-mode equations. We derive the exact expressions of the overlap integrals and coupling coefficients by utilizing the addition theorems of Bessel functions.  We then generalize the discussion to an infinite waveguide array and rigorously characterize the dispersion relation and continuum with the help of theories in harmonic analysis.  In the second part of the paper, we give a strict proof of the existence of BICs in the aforementioned waveguide system with two additional identical vertical waveguides aligned symmetrically above and below the horizontal waveguide array. We further numerically demonstrate the transition from a perfect BIC to a leaky mode by introducing a symmetry-breaking refractive index perturbation and quantitatively analyze the resulting radiation losses. This work gives a comprehensive study of symmetry-protected BICs and provides an efficient and precise computational model for designing such BICs devices.
\end{abstract}




\end{frontmatter}




\section{Introduction}
\label{sec1}
In wave systems, states are conventionally classified into bound states and continuum states. Bound states are confined within certain structures and possess discrete energy levels. In contrast, continuum states are extended, radiate energy away, and form a continuous energy spectrum. However, in 1929, von Neumann and Wigner ~\cite{neumann1929merkwurdige,von1993merkwurdige} proposed a counterintuitive class of solutions to the wave equation: bound states in the continuum (BICs).
BICs represent a counter-intuitive wave phenomenon where a localized eigenstate coexists with a continuous spectrum of radiating waves without leaking energy~\cite{hsu2016bound}. Originally proposed  in quantum mechanics, BICs have found a fertile ground for experimental realization in photonics, particularly in waveguide arrays where the paraxial diffraction of light mimics the temporal evolution of a quantum wave packet. Plotnik et al.~\cite{Plotnik2011OpticalBIC} experimentally demonstrated symmetry-protected BICs in such systems, utilizing an antisymmetric mode decoupled from a symmetric continuum.

In optical systems, BICs usually occur due to symmetry mismatch between the bound state and all available radiation modes. Such BICs, often named symmetry-protected BICs, have been experimentally observed in waveguide arrays where an antisymmetric localized mode remains isolated within a continuum of symmetric extended states. While the physical mechanism of symmetry protection is well-understood conceptually~\cite{hsu2016bound}, accurate mathematical modeling of such systems requires careful treatment of the waveguide interactions. Standard theoretical approaches often rely on the tight-binding approximation or orthogonal coupled-mode theory~\cite{yang2024programmable, petravcek2022bound, xia2015supermodes, xia2011supermodes}, which may overlook the non-vanishing overlap between the modes of adjacent waveguides and thus lack full mathematical rigor.  For densely packed waveguide arrays, the non-orthogonality of the eigenmodes becomes significant. Although some studies based on nonorthogonal coupled-mode equations exist, their scope of applicability and availability of concise formulas for waveguide arrays remain limited~\cite{suh2004temporal, pinske2022symmetry, huang1994coupled, haus2003coupled}. Consequently, a rigorous formulation based on NCME would be beneficial for the accurate computation of dispersion relations and coupling coefficients.

In this work, we provide a comprehensive mathematical derivation and numerical verification of BICs in a coupled waveguide system consisting of an infinite horizontal array and two vertical defects. Our approach distinguishes itself by deriving exact analytical representations for the entries of the coupling matrices ($\bm{S}$ and $\bm{K}$). By employing the addition theorems for Bessel functions and modified Bessel functions, we evaluate the overlap integrals explicitly, avoiding the errors associated with numerical integration in unbounded domains. Furthermore, we treat the infinite waveguide array using the theory of bounded linear operators on the Hilbert space $\ell^2(\mathbb{Z})$. Within the framework of Banach algebras, we utilize the properties of convolution operators to characterize the system's band structure. By invoking Wiener's lemma on the algebra of Fourier series, we define and derive the continuum rigorously and verify the embedding of the discrete bound state eigenvalues. In the final part of the paper, we break the symmetry of the waveguide system by introducing a perturbation to the refractive index in the vertical waveguides. Numerical experiments confirm that, once the symmetry is broken, BICs no longer exist and the energy dissipates to infinity through the horizontal waveguide system.

The structure of this paper is organized as follows. In Section 2, we formulate the problem using the paraxial wave equation and introduce the single-mode condition for the step-index fibers. Section 3 is dedicated to the derivation of the Nonorthogonal Coupled-Mode Equations (NCME). We prove that the coupling matrices are Toeplitz and symmetric, and we provide their explicit analytical formulas. In Section 4, we extend the NCME to the infinite waveguide system and calculate the dispersion relation and the  continuum by analyzing the Fourier symbol of the convolution kernel. In Section 5, we demonstrate the occurrence of the BIC by verifying that the antisymmetric vertical mode is decoupled from the horizontal continuum. Finally, in Section 6, we introduce a refractive index perturbation to break the symmetry, mathematically modeling the transition from a perfect BIC to a leaky mode and verifying the results through numerical simulation via a power-conserving Crank-Nicolson scheme.

\section{Formulation of the problem}
In this work, we focus on modeling the formation and evolution of a symmetry-protected bound state in the continuum (BIC), which has been observed in the experimental setup \cite{Plotnik2011OpticalBIC}. The system consists of an array of 51 cylindrical waveguides extending along $z$-direction and evenly spaced along $x$-direction, labeled from $-25$ to $25$ with the central waveguide as $0$. Two additional vertical waveguides are positioned above and below the central horizontal waveguide at coordinate $(0,0)$ along the $y$-direction, denoted as $+$ and $-$, respectively. In \cite{Plotnik2011OpticalBIC}, the refractive index profile of each waveguide follows a sixth-order Gaussian distribution, and for simplicity, the refractive index distribution is approximated as piecewise constant and independent of $z$, that is, the waveguides are regular. We assume that light propagates along the $z$-direction and that each waveguide supports a single guided mode.

In the transverse plane, the centers of the horizontal waveguides are located at $(mD_h, 0)$, $m\in [-25,25]\cap\mathbb{Z}=:\mathbb{Z}^{(25)}$ while the centers of the vertical waveguides are located at
$(0,\,\pm D_v).$
Here $D_h$ and $D_v$ denote the center-to-center spacings of the horizontal and vertical waveguides, respectively.
Let $a$ be the radius of each waveguide, and assume  $a<\min(D_h, D_v)$. The boundary $\mathbb{B}$ of waveguide array system is the union of the every waveguide boundary, namely,
$$\mathbb{B}=\mathbb{B}_h\cup\mathbb{B}_{v}=\bigcup\limits_{m\in\mathbb{Z}^{(25)}} \mathbb{B}_m \cup \mathbb{B}_v,
$$
where $\mathbb{B}_m=\partial B_a(mD_h,0)\times\mathbb{R}_{\geqslant 0}$ 
denotes the boundary of the $m$-th horizontal waveguide and $\mathbb{B}_v=(\partial B_a(0,D_v)\cup \partial B_a(0,-D_v))\times\mathbb{R}_{\geqslant  0}$ represents the boundaries of the two vertical waveguides.

Denote $n(x,y,z) = n_0 + \Delta n(x,y)$ as the refractive index profile in $\mathbb{R}^3$, where $\Delta n(x,y) = \Delta n_{+}(x,y)+\Delta n_{-}(x,y)+\sum\limits_{m\in\mathbb{Z}^{(25)}} \Delta n_m(x,y)$.
The components $\Delta n_{+}(x,y), \Delta n_{-}(x,y)$ and $\Delta n_m(x,y)$ are defined by
\begin{equation*}
\Delta n_\pm(x,y) =
\left\{
\begin{aligned}
\Delta n, && (x,y) \in B_a(0,\pm D_v)\\[3pt]
0, && \text{otherwise},
\end{aligned}
\right.
\quad
\Delta n_m(x,y) =
\left\{
\begin{aligned}
\Delta n, && (x,y) \in B_a(mD_h, 0)\\[3pt]
0, && \text{otherwise}.
\end{aligned}
\right.
\end{equation*}
We require $\Delta n/n_0 \ll 1$, i.e. the waveguides are \emph{weakly guided}.

\begin{remark}\label{rmk:1}
Experiment in \cite{Plotnik2011OpticalBIC} used $a = 3.32\,\mu\text{m}, \Delta n = 8 \times 10^{-4}, n_0 = 1.45, D_h = 20\,\mu\text{m}, D_v = 15\,\mu\text{m}$.
\end{remark}
\begin{figure}[htbp]
   \centering
   \includegraphics[width=0.8\textwidth]{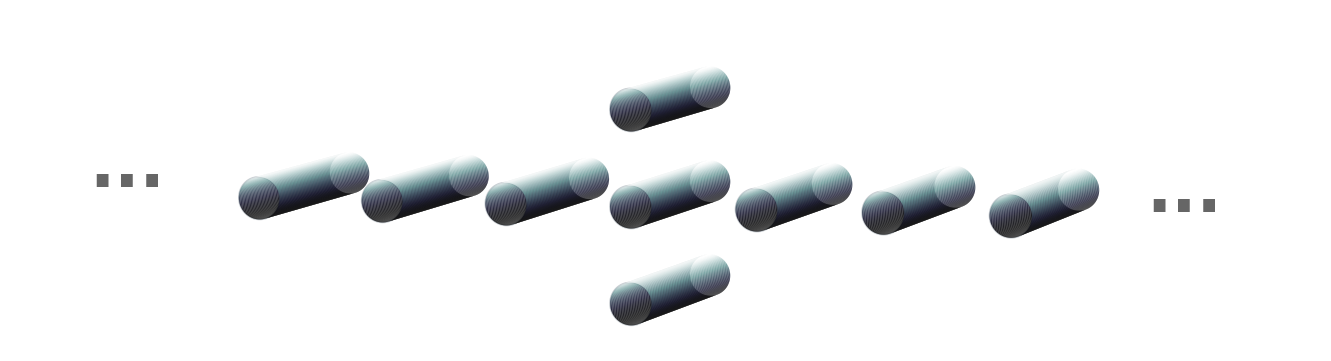} 
   \caption{Waveguide array}
   \label{Waveguide array}
\end{figure}

\subsection{Field Representation and Paraxial Wave Equation}

We assume the electromagnetic wave propagating in the system as a monochromatic time-harmonic wave with frequency $\omega$, then each Cartesian component of the electric field $E_p(x,y,z)$, $p=x,y,z$ satisfies the scalar Helmholtz equation
$$
\nabla^2 E_p + \left(\frac{\omega}{c}\right)^2 n^2(x,y,z) E_p = 0.
$$
The magnetic field $\bm{H}$ can be derived similarly; thus, we focus exclusively on the electric field. Without loss of generality, the electric field is assumed to be polarized along the $x$-axis for convenience and symmetry in the analysis. In the weakly guiding limit ($\Delta n/n_0\ll 1$), the longitudinal field components are of order $\Delta n/n_0$ and thus negligible compared to the transverse components. Therefore $E_z\approx 0$, and we only need to calculate $E_x$, which will be denoted as $E$ in the rest of paper.  Weakly guiding limit implies that $E_x$ is slowly varying, defined as follows (cf. Chapter 5.1 of \cite{Lifante2003IntegratedPhotonicsFundamentals}):
\begin{definition}
The amplitude $\psi(x,y,z)\in C^2(\mathbb{R}^3)$ is \emph{slowly varying} if
$$
\vert \partial_z^2\psi\vert\ll k\vert \partial_z\psi\vert,
$$
where $k=k_0n_0=\dfrac{2\pi}{\lambda}n_0$ is the background wavenumber.
\end{definition}
Since the dominant propagation is along $+z$, the slowly varying $E$ is in the form of
$$
E(x,y,z) = \psi(x,y,z) \, e^{\ii kz},
$$
where the function $\psi(x,y,z)$ is called the \emph{envelope} varies slowly relative to the carrier $e^{\ii kz}$. Substituting this form into the scalar Helmholtz equation and neglecting the $\partial_z^2 \psi$ term due to the slowly varying nature, we achieve the the governing equations to be solved.
\begin{definition}[Solution to the paraxial wave equation]
Let $\Omega = \mathbb{R}^2$ denote the transverse plane, and $z\in \mathbb{R}_{\geqslant 0}$ the propagation direction. A function
$$\begin{aligned}\psi:\Omega\times \mathbb{R}_{\geqslant 0} &\to \mathbb{C}\\
(x,y,z)&\mapsto \psi(x,y,z)
\end{aligned}$$
is called a \emph{solution} of the waveguide array system,
if $\psi \in C^2(\Omega \times \mathbb{R}_{\geqslant 0}\setminus\mathbb{B})\cap C^1(\mathbb{B})$, and for any fixed $z$, $\psi\in L^2(\Omega)$ satisfying
\begin{equation}\label{eq:defsol}
\left\{\begin{aligned}&\left(i\partial_z+\dfrac{1}{2k}(\partial_x^2+\partial_y^2)+\dfrac{k\Delta n(x,y)}{n_0}\right)\psi=0,\quad (x,y,z)\in \mathbb{R}^3\setminus\mathbb{B}, \\
&\psi(x, y, z)\vert_- = \psi(x, y, z)\vert_+, \quad (x,y,z)\in \mathbb{B},\\&\left.\frac{\partial \psi}{\partial n}\right|_-(x, y, z) = \left.\frac{\partial \psi}{\partial n}\right|_+(x, y, z),\quad(x,y,z)\in \mathbb{B}.
\end{aligned}\right.
\end{equation}
\end{definition}

Since the waveguide system considered in this paper is invariant along $z$, the envelope could be written as
$$
\psi(x,y,z) = \Phi(x,y) \, e^{\ii (\delta k) z}.
$$
$\Phi(x,y)$ denotes the \emph{transverse mode profile}, and $\delta k$ is the \emph{longitudinal wavenumber shift} from $k$.

\subsection{Single guiding mode condition}
For a single isolated uniform waveguide, with $n(x,y)$ independent of $z$, we denote this shift specifically as $\beta_0$. Then when the whole system consists of one single waveguide, namely the central waveguide, the envelope $\psi_0$ can be written as
$$
\psi_0(x,y,z) = \Phi_0(x,y)e^{i\beta_0 z}.
$$

From \eqref{eq:defsol}, $\Phi_0(x,y)$ satisfies
\begin{equation}\label{eq:defsol1}
\left\{\begin{aligned}&\left(\dfrac{1}{2k}(\partial_x^2+\partial_y^2)+\dfrac{k\Delta n_0(x,y)}{n_0}\right)\Phi (x,y)-\beta_0\Phi(x,y)=0,\quad (x,y)\in \Omega\setminus \partial B_a(0,0),\\
&\Phi(x,y)\vert_-=\Phi(x,y)\vert_+,\quad (x,y)\in \partial B_a(0,0),\\
&\left.\dfrac{\partial\Phi}{\partial n}(x,y)\right|_-=\left.\dfrac{\partial\Phi}{\partial n}(x,y)\right|_+,\quad (x,y)\in \partial B_a(0,0).
\end{aligned}\right.
\end{equation}
Solution to \eqref{eq:defsol1} is called \emph{transverse mode}.

Define \begin{equation}\label{formulaLambdaGamma}
\Lambda:=\sqrt{\dfrac{2k^2\Delta n}{n_0}-2k\beta_0}=\sqrt{\dfrac{8\pi^2n_0\Delta n}{\lambda^2}-\dfrac{4\pi n_0\beta_0}{\lambda}},\quad \Gamma:=\sqrt{2k\beta_0}=\sqrt{\dfrac{4\pi n_0\beta_0}{\lambda}}.
\end{equation}

The first equation in \eqref{eq:defsol1} can be reduced to a radial Sturm–Liouville problem
\begin{align}\label{2.13}
(\partial_x^2+\partial_y^2)\Phi(x,y)+\Lambda^2\Phi(x,y)&=0,\quad (x,y)\in B_{a}(0)\\
\label{2.14}(\partial_x^2+\partial_y^2)\Phi(x,y)-\Gamma^2\Phi(x,y)&=0,\quad (x,y)\in\Omega\setminus \overline{B_{a}(0)}.
\end{align}
Since the waveguide is radially symmetric, we could assume that $$\Phi(x,y)=\phi(r)R(\theta).$$
Then \eqref{eq:defsol1} becomes
\begin{numcases}{}
r^2\dfrac{\mathrm{d}^2\phi}{\mathrm{d}r^2}
 + r\dfrac{\mathrm{d}\phi}{\mathrm{d}r}
 + (r^2\Lambda^2-\zeta)\phi = 0,  \quad r\in [0,a),\label{eq-Besseleq} \\
r^2\dfrac{\mathrm{d}^2\phi}{\mathrm{d}r^2}
 + r\dfrac{\mathrm{d}\phi}{\mathrm{d}r}
 - (r^2\Gamma^2+\zeta)\phi = 0, \quad  r\in (a,+\infty),\label{eq-Besseleq-2} \\
 \dfrac{\mathrm{d^2}}{\mathrm{d}\theta^2}R(\theta)+\zeta R(\theta)=0,\\
\phi(r) = \phi(r)|_-=\phi(r)|_+, \quad r=a, \label{eq-3.12}\\
\phi'(r) = \phi'(r)|_-=\phi'(r)|_+, \quad r=a, \label{eq-3.13}\\
R(\theta)=R(\theta+2\pi).
\end{numcases}
It is immediate that $R(\theta)=e^{\ii m\theta}$ and the eigenvalues are $\zeta = m^2$ for $m\in\mathbb{Z}$. By substituting $x=\Lambda r$, \eqref{eq-Besseleq} then becomes
$$
x^2\dfrac{\mathrm{d}^2\phi}{\mathrm{d}x^2}
 + x\dfrac{\mathrm{d}\phi}{\mathrm{d}x}
 + (x^2-m^2)\phi = 0,
 $$
which is a Bessel equation of order $m$.

Similarly, the equation \eqref{eq-Besseleq-2} can be reduced to  a modified Bessel equation of order $m$.
\begin{remark}\label{rmk:2}
This case matches the standard fiber theory, and  it leads to the single mode condition:
$$\lambda>\sqrt{8\pi^2a^2n_0\Delta n/j^2_{0,1}},$$
where $j_{0,1}$ is the first zero of the first kind Bessel function $J_{0}$. According to the single-mode condition, the guiding wavelength must satisfy $\lambda > 0.418~\mu\mathrm{m}$. In the following, we take the laser wavelength to be $800~\mathrm{nm}$, which is typically used in experiments.
\end{remark}

Let $J_l$ be the Bessel function of the first kind of order $l$, $I_m$ be the modified Bessel function of first kind of order $m$ and $K_n$ be the modified Bessel function of second kind of order $n$. As in Chapter 3.2 of \cite{Okamoto2006Fundamentals}, the \emph{fundamental transverse mode} $\phi(r)$ takes the form of
\begin{equation}\label{eq-3.16}
\phi(r)=\left\{
\begin{aligned}
AJ_0(\Lambda r)&,\quad r\in(0,a),\\
BK_0(\Gamma r)&,\quad r\in(a,\infty),
\end{aligned}\right.
\end{equation}
We assume that $A$ and $B$ takes the value calculated in Proposition \ref{prop:valueab} such that the mode is normalized. From boundary conditions \eqref{eq-3.12} and \eqref{eq-3.13} $\Lambda$ and $\Gamma$ should satisfy
\begin{equation}\label{governbeta}
\dfrac{\Lambda J_1(\Lambda a)}{J_0(\Lambda a)} = \dfrac{\Gamma K_1(\Gamma a)}{K_0(\Gamma a)}.
\end{equation}

\begin{remark}\label{rmk:3}
We can calculate $\beta_0$ from the implicit equation \eqref{governbeta} numerically. With parameters from Remark \ref{rmk:1}, $\beta_0=808.07\text{m}^{-1}$.
\end{remark}

\subsection{Properties of the fundamental transverse mode}
\begin{proposition}\label{prop-3.1}The mode $\phi(r)$\eqref{eq-3.16} is positive, decreasing and bounded for $r\in [0,+\infty)$.
 \end{proposition}
 \begin{proof}Since the point $r=a$ doesn't reach the first zero of the function $J_0(\Lambda x)$, $J(0)=1$ and the function $J_0(\Lambda x)$ is decreasing on $[0,a]$, the function $\phi(r)$ is positive on $[0,a]$. And $\phi(r)=BK_0(\Gamma r)$ on $[a,+\infty)$, which is positive and decreasing. Hence $0<\phi(r)\leqslant \phi(0)=A$.
 \end{proof}

\begin{proposition} The characteristic equation
\begin{equation}\label{goverbeta}\dfrac{\Lambda J_1(\Lambda a)}{J_0(\Lambda a)} = \dfrac{\Gamma K_1(\Gamma a)}{K_0(\Gamma a)},\end{equation}
is a necessary condition such that transverse mode function $\phi(x,y)$ is continuously differentiable. Moreover, the transverse mode function $\phi(x,y)$ is infinitely differentiable except $r=a$. In fact, it is not twice differentiable at $r=a$.
 \end{proposition}
 \begin{proof}
 Since $\phi(r)$ is continuous at $r=a$, it follows that $\phi(a^-) = \phi(a^+)$, namely, $A J_0(\Lambda a) = B K_0(\Gamma a).$ The derivative of $\phi(r)$ at $r=a$ is also continuous, which implies $A \Lambda J_1(\Lambda a) = B \Gamma K_1(\Gamma a).$ Note that $J_0(\Lambda a)$ is positive by Proposition \ref{prop-3.1}. Therefore, the identity \eqref{goverbeta} is valid as desired.
 However, the second derivative of $\phi(r)$ at $r=a$ cannot be continuous, because if $\phi''({a^-})=\phi''({a^+})$ were true, it would imply $\Gamma^2+\Lambda^2=0,$
which is a contradiction.
 \end{proof}

\begin{proposition}\label{prop:valueab} If $\phi_0(x,y)=\phi(x,y)$ is normalized, that is, $$\iint_{\Omega}\vert\phi_0(x,y)\vert^2\mathrm{d}x\mathrm{d}y=1,$$ then
\begin{equation}\label{formulaAB}A = \left[ \pi a^2 \left( J^2_1(\Lambda a) + J^2_0(\Lambda a) \right) + \pi a^2 \left(\dfrac{J_0(\Lambda a)}{K_0(\Gamma a)}\right)^2\left(K^2_1(\Gamma a)-K^2_0(\Gamma a)\right)\right]^{-1/2}, \quad B = A \frac{J_0(\Lambda a)}{K_0(\Gamma a)}.\end{equation}
\end{proposition}
\begin{proof} By  \ref{Besselint} and the equation \eqref{governbeta}, it follows the formulas for $A$ and $B$ immediately.
\end{proof}
Then we always assume that the modes are normalized.

\section{Nonorthogonal Coupled-Mode Equations}\label{sec:3}

\subsection{Coupled-mode equation}\label{subsec-3.5}
Assume that the  envelope of the waveguide array is a superposition of these transverse modes with $z$-dependent amplitudes:
\begin{equation}\label{eq:totalenvelope}
\psi(x,y,z)=\sum\limits_{m\in\mathbb{Z}^{(25)}}c_m(z)\phi_m(x,y),
\end{equation}
where $\phi_m(x,y)=\Phi_0(x-mD_h,y)$ denotes the mode of waveguide $m$, and $c_m(z)\in\mathbb{C}$ are the coefficients.

Define operators \begin{equation}\label{def_H}\widehat{H}_m=\left(\dfrac{1}{2k}(\partial_x^2+\partial_y^2)+\dfrac{k\Delta n_m(x,y)}{n_0}\right)\quad \text{and}\quad \widehat{H}=\left(\dfrac{1}{2k}(\partial_x^2+\partial_y^2)+\dfrac{k\Delta n(x,y)}{n_0}\right).\end{equation}
    Notice that from \eqref{eq:defsol1}, $\beta_0$ is the eigenvalue of the operator $\widehat{H}_m$, i.e. $\widehat{H}_m\phi_m(x,y)=\beta_0\phi_m(x,y)$.
Substituting \eqref{eq:totalenvelope} into the paraxial wave equation \eqref{eq:defsol} yields
\begin{equation}
\sum_m\ii\phi_m(x,y)\partial_z c_m(z)+\sum_m c_m(z)\widehat{H}\phi_m(x,y)=0.
\end{equation}
Multiply by the mode function $\phi_n(x,y)$ and integrate over $\Omega$:
\begin{equation}\label{eq-3.25}
\ii\sum\limits_m\partial_z c_m(z)\iint_{\Omega}\phi_n(x,y)\phi_m(x,y)\mathrm{d}x\mathrm{d}y+\sum\limits_{m}c_m(z)\iint_{\Omega}\phi_n(x,y)\widehat{H}\phi_m(x,y)\mathrm{d}x\mathrm{d}y=0.\end{equation}

To simplify the first term, we define $\bm{S}=(S_{ij})$ and $\bm{K}=(K_{ij})$ by
$$
S_{ij}:=\iint_{\Omega}\phi_i(x,y)\phi_j(x,y)\mathrm{d}x\mathrm{d}y, \qquad K_{ij} := \iint_{\Omega}\phi_i(x,y)\widehat{H}\phi_j(x,y)\mathrm{d}x\mathrm{d}y.
$$
\eqref{eq-3.25} is then equivalent to
$$
    \ii\sum\limits_mS_{nm}\partial_z c_m(z) + \sum\limits_{m}K_{nm}c_m(z) = 0.
$$
Define $\bm{C}(z):=(c_m(z))_{m\in\mathbb{Z}^{(25)}}\in \mathbb{C}^{51}$, the equation \eqref{eq-3.25} can be reformulated in the following compact form
\begin{equation}\label{eq:ncme}
\ii\bm{S}\dfrac{\mathrm{d}}{\mathrm{d}z}\bm{C}(z)+\bm{K}\bm{C}(z)=\bm{0}.
\end{equation}
This equation is called the non-orthogonal coupled-mode equation (NCME) for a finite waveguide system. Later we will also generalize \eqref{eq:ncme} to infinite waveguide system case. In the infinite case, the index $m$ takes values in $\mathbb{Z}$.

\subsection{Exact representation for entries of coupled matrices}

In this part we are dedicated to deriving explicit expressions of $S_{nm}$ and $K_{nm}$. It is worth noting that the formula in the following for $\bm{K}$ is valid only in the infinite case, whereas that for $\bm{S}$ holds in both finite and infinite waveguide systems. Nevertheless, for the finite system, $\bm{K}$ may still be evaluated following the same procedure.\\
Firstly we decompose $K_{nm}$ as
$$
\begin{aligned}
K_{nm}&=\iint_{\Omega}\phi_n(x,y)\widehat{H}\phi_m(x,y)\mathrm{d}x\mathrm{d}y =\iint_{\Omega}\phi_n(x,y){\left(\widehat{H}_m\phi_m(x,y)+\sum\limits_{l\neq m}\dfrac{k\Delta n_l(x,y)}{n_0}\phi_m(x,y)\right)}\mathrm{d}x\mathrm{d}y\\
&=\beta_0S_{nm}+\iint_{\Omega}\phi_n(x,y)\sum\limits_{l\neq m}\dfrac{k\Delta n_l}{n_0}\phi_m(x,y)\mathrm{d}x \mathrm{d}y.\end{aligned}$$
Define $\bm{\kappa} = (\kappa_{ij})$ by
\begin{equation}\label{defkappa}
\kappa_{ij}:=\iint_{\Omega}\phi_i(x,y)\sum\limits_{l\neq j}\dfrac{k\Delta n_l(x,y)}{n_0}\phi_j(x,y)\mathrm{d}x \mathrm{d}y.
\end{equation}
Then
\begin{equation}\label{eq:defskappa}
K_{nm}=\beta_0S_{nm}+\kappa_{nm}.
\end{equation}

\begin{lemma}\label{Hselfadjoint}The operator $\widehat{H}$ is self-adjoint on the Sobolev space $H^2(\Omega)$.(See Reed~\cite[p.~54, Theorem~IX.27]{reed1972methods} or Hislop~\cite[p.~2]{hislop2012fundamentals} or Chipot~\cite[p.~471]{chipot2005handbook})
\end{lemma}
 \begin{proposition}\label{propSKsym}The matrices $\bm{S}$, $\bm{K}$, and $\bm{\kappa}$ are symmetric. Moreover, the entries of $\bm{S}$    depend only on the index difference $\lvert i-j\rvert$, so $\bm{S}$ is a Toeplitz matrix. These properties of $\bm{S}$, $\bm{K}$, and $\bm{\kappa}$ hold for both the finite and infinite cases. However, the Toeplitz property (dependence solely on $\lvert i - j \rvert$) for $\bm{K}$ and $\bm{\kappa}$ holds strictly only in the infinite-waveguide case.
\end{proposition}
\begin{proof}By the definition of $\bm{S}$, its symmetry follows directly. Applying Lemma~\ref{Hselfadjoint}, it follows that
$$K_{ij}=\iint_{\Omega}\phi_i(x,y)\widehat{H}\phi_j(x,y)\mathrm{d}x\mathrm{d}y=\iint_{\Omega}\phi_j(x,y)\widehat{H}\phi_i(x,y)\mathrm{d}x\mathrm{d}y=K_{ji}.$$ According to the relation \eqref{eq:defskappa}, one can also discern that $\kappa_{ij}=\kappa_{ji}$.\\
Using the change of variables $x=u+iD_h$ and $y=v$, we have
$$S_{ij}=\iint_{\Omega}\phi_0(u,v)\phi_{j-i}(u,v)\mathrm{d}u\mathrm{d}v,$$
and
$$\begin{aligned}\kappa_{ij}&=\dfrac{k\Delta n}{n_0}\iint_{\bigcup\limits_{l\neq j}B_a(l D_h,0)}\phi_i(x,y)\phi_j(x,y)\mathrm{d}x \mathrm{d}y&=\dfrac{k\Delta n}{n_0}\iint_{\bigcup\limits_{l\neq j-i}B_a(l D_h,0)}\phi_0(u,v)\phi_{j-i}(u,v)\mathrm{d}u \mathrm{d}v.
\end{aligned}$$
Therefore, the entries of $\bm{S}$ only depend on $\vert i-j \vert$.
However, the substitution we used is valid only in the infinite case for $\kappa_{ij}$. Hence, the entries of $\bm{\kappa}$ and $\bm{K}$ only depend on $\vert i-j \vert$ in the infinite case.
\end{proof}
From now on we use the notation $S_{\xi}$, $K_{\xi}$ and $\kappa_{\xi}$ to represent the entries of three matrices respectively.

\begin{theorem}The entries of $\bm{S}$ are given by \begin{equation}\label{formulaSh}\begin{aligned}
S_0=1,\quad 
S_\xi =&\dfrac{4a\pi ABK_0(\Gamma d)}{\Gamma^2+\Lambda^2}\left[\Lambda I_0(\Gamma a)J_1(\Lambda a)+\Gamma J_0(\Lambda a)I_1(\Gamma a)\right]
+\dfrac{\pi B^2 d K_1(\Gamma d)}{\Gamma}\\
&-2\pi B^2 a^2K_0(\Gamma d)\left[I_0(\Gamma a)K_0(\Gamma a)+I_1(\Gamma a)K_1(\Gamma a)\right],\end{aligned}\end{equation}
where $d=\xi D_h$, $\xi \in\mathbb{Z}_{>0}$ is the distance between two waveguides.\end{theorem}
\begin{proof}Since the functions $\phi_k(x,y)$ and $\phi_0(x,y)$ are piecewise, we separate the plane $\Omega$ into three subsets:
$\Omega_1=B_a(0,0),\Omega_2=B_a(d,0)$ and $\Omega_3=\Omega\setminus(\Omega_1\cup\Omega_2)$. Using polar coordinates for each subset, one can obtain
$$\begin{aligned}S_{\xi}&=\iint_{\Omega_1}\phi_0(x,y)\phi_\xi(x,y)\mathrm{d}x\mathrm{d}y+\iint_{\Omega_2}\phi_0(x,y)\phi_\xi (x,y)\mathrm{d}x\mathrm{d}y+\iint_{\Omega_3}\phi_0(x,y)\phi_\xi(x,y)\mathrm{d}x\mathrm{d}y\\&=:I_1+I_2+I_3\end{aligned}$$
where $$I_1 =I_2 = A B \int_0^{2\pi} \mathrm{d}\theta\int_0^a J_0(\Lambda r) K_0(\Gamma \sqrt{r^2 + d^2 - 2 d r \cos \theta}) r \mathrm{d}r$$
and $$I_3=B^2\iint_{\Omega_3}  K_0(\Gamma \sqrt{x^2 + y^2}) K_0(\Gamma \sqrt{(x - d)^2 + y^2}) \mathrm{d}x \mathrm{d}y.$$
To evaluate the integral $I_1$ and $I_2$, one can apply the addition theorem for $K_0$ (Lemma~\ref{Additiontheorem}) and get
$$K_0(\Gamma \sqrt{r^2 + d^2 - 2 d r \cos \theta}) = \sum_{m=-\infty}^{\infty} I_m(\Gamma r) K_m(\Gamma d) \cos(m \theta)$$
for $r\leqslant a<d$. Therefore, the angular integral gives
$$
\int_0^{2\pi} K_0(\Gamma \sqrt{r^2 + d^2 - 2 d r \cos \theta}) \, d\theta = 2\pi I_0(\Gamma r) K_0(\Gamma d).
$$
By applying identity \eqref{eq:bessel4}, we conclude that
\begin{equation}\label{eq:i1}
I_1 = 2\pi A B K_0(\Gamma d) \cdot \frac{a}{\Lambda^2 + \Gamma^2} \left[ \Lambda I_0(\Gamma a) J_1(\Lambda a) + \Gamma J_0(\Lambda a) I_1(\Gamma a) \right].
\end{equation}
Evaluating the integral over $\Omega_3$ directly is challenging due to its geometry. Instead, we employ a domain decomposition technique:
$$I_3=I_3^{total}-2I_3^{disk}$$
where $$I_3^{total}= B^2 \iint_{\Omega} K_0(\Gamma \sqrt{x^2 + y^2}) K_0(\Gamma \sqrt{(x - d)^2 + y^2}) \mathrm{d}x\mathrm{d}y,$$ and $$I_3^{disk}= B^2 \iint_{B_a(0,0)} K_0(\Gamma \sqrt{x^2 + y^2}) K_0(\Gamma \sqrt{(x - d)^2 + y^2}) \mathrm{d}x\mathrm{d}y.$$
In polar coordinates,
$$I_3^{total} = B^2 \int_0^{2\pi} \mathrm{d}\theta\int_0^\infty K_0(\Gamma r) K_0(\Gamma \sqrt{r^2 + d^2 - 2 r d \cos \theta}) r\mathrm{d}r .$$
Applying the addition theorem for $K_0$ again yields
$$\begin{aligned}I_3^{total} &= B^2 \int_0^{2\pi}\mathrm{d}\theta  \left(\int_0^d+\int_d^\infty\right) K_0(\Gamma r) \sum_{m} I_m(\Gamma r) K_m(\Gamma d) \cos(m \theta) r \mathrm{d}r \\
&= 2\pi B^2 \left[ K_0(\Gamma d) \int_0^d K_0(\Gamma r) I_0(\Gamma r) r \mathrm{d}r + I_0(\Gamma d) \int_d^\infty K_0^2(\Gamma r) r \mathrm{d}r\right].\end{aligned}$$
Then from \ref{Besselint}
\begin{equation}\label{eq:i3total}
I_3^{total}=\pi B^2d^2K_1(\Gamma d)\left[K_0(\Gamma d)I_1(\Gamma d)+I_0(\Gamma d)K_1(\Gamma d)\right].
\end{equation}
Similarly,
\begin{equation}\label{eq:i3disk}
\begin{aligned}
I_3^{disk}&=2\pi B^2K_0(\Gamma d)\int_0^aK_0(\Gamma r)I_0(\Gamma r)r\mathrm{d}r=\pi B^2 a^2K_0(\Gamma d)\left[K_0(\Gamma a)I_0(\Gamma a)+I_1(\Gamma a)K_1(\Gamma a)\right].
\end{aligned}
\end{equation}
By plugging \eqref{eq:i1}, \eqref{eq:i3total} and \eqref{eq:i3disk} into equation $S_\xi = 2I_1+ I_3^{total} - 2I_3^{disk}$ we arrive at the desired formula after a slight simplification using the known Wronskian identity for modified Bessel functions: $$I_0(z) K_1(z) + I_1(z) K_0(z) = \frac{1}{z}.$$\end{proof}

\begin{lemma}\label{lem-3.14}Let $A_1$, $A_2$ be two positive constants. Then the sequence $\{T_m\}_{m\in\mathbb{Z}}$ defined by
$$
T_m:=K_m(A_1)K_m(A_2)\int_0^aI^2_m(\Gamma r)r\;\mathrm{d}r
$$
satisfies
$$T_m\leqslant C_{\Gamma, a}\;\dfrac{q^m}{m^2}\leqslant C_{\Gamma, a}\; q^m,\quad m\in\mathbb{Z}_{>0},$$
where $C_{\Gamma, a}$ is a constant depending on $\Gamma$ and $a$, and $q=\dfrac{\Gamma^2 a^2}{A_1A_2}$. Therefore, if $q<1$, then the series $\sum_{m=1}^\infty T_m$ is absolutely convergent.
\end{lemma}
\begin{proof}
By Lemma \ref{lem-3.8} and Lemma \ref{lem-3.9}, it follows that
$$K_m(A_1)K_m(A_2)\leqslant \dfrac{1}{4}\bm{\Gamma}^2(m)\left(\dfrac{4}{A_1A_2}\right)^m\quad \text{ and }\quad I^2_m(\Gamma r)\leqslant \dfrac{(\Gamma r)^{2m}e^{2\Gamma r}}{2^{2m}(m!)^2}\leqslant \dfrac{(\Gamma a)^{2m}e^{2\Gamma a}}{2^{2m}(m!)^2},$$
where $\bm{\Gamma}(m)$ stands for Gamma function.

Therefore, $$\int_0^aI^2_m(\Gamma r)r\;\mathrm{d}r\leqslant \dfrac{e^{2\Gamma a}}{(m!)^2}\left(\dfrac{\Gamma a}{2}\right)^{2m}\dfrac{a^2}{2}.$$
Then, $$T_m\leqslant \dfrac{e^{2\Gamma a}a^2}{8} \dfrac{\bm{\Gamma}^2(m)}{(m!)^2}\left(\dfrac{(\Gamma a)^2}{A_1A_2}\right)^m= \dfrac{e^{2\Gamma a}a^2}{8}\dfrac{1}{m^2}\left(\dfrac{(\Gamma a)^2}{A_1A_2}\right)^m.$$
Let $C_{\Gamma, a}:=\dfrac{e^{2\Gamma a}a^2}{8}$ and $q:=\dfrac{(\Gamma a)^2}{A_1A_2}$, then the conclusion appears as desired.
\end{proof}

Before calculating $\kappa_{s}$ for infinite case, we define two series and one integral
\begin{equation}\label{eq:auxint1}
\begin{aligned}
\varsigma_{l,s}:=&K_0(\Gamma \vert l\vert D_h) K_0(\Gamma \vert l-s\vert D_h) \int_0^a I^2_0(\Gamma r) r \, dr\\&+2\sum_{m=1}^\infty \left[K_m(\Gamma \vert l\vert D_h) K_m(\Gamma \vert l-s\vert D_h) \int_0^a I^2_m(\Gamma r) r \, dr\right],\quad l(l-s)>0,\\
\zeta_{l,s}:=&K_0(\Gamma \vert l\vert D_h) K_0(\Gamma\vert l-s\vert D_h) \int_0^a I^2_0(\Gamma r) r \, dr\\&+2\sum_{m=1}^\infty \left[(-1)^m K_m(\Gamma \vert l\vert D_h) K_m(\Gamma\vert l-s\vert D_h) \int_0^a I^2_m(\Gamma r) r \, dr\right],\quad 0<l<s,\\
\tau_s:=&K_0(\Gamma s D_h)\int_0^a J_0(\Lambda r)I_0(\Gamma r)r\mathrm{d}r,\quad l=0, s> 0.
\end{aligned}
\end{equation}
Using Lemma \ref{lem-3.14} and \ref{Besselint}, the series
$$
\sum_{m=1}^\infty K_m(\Gamma\vert l\vert  D_h) K_m(\Gamma \vert l-s\vert D_h) \int_0^a I_m(\Gamma r)^2 r \, dr
$$
converges since $q=\dfrac{a^2}{\vert l(l-s)\vert D^2_h}<1$, therefore $\varsigma_{l,s}$ and $\zeta_{l,s}$ are well-defined. The integrals $\int_0^a I^2_m(\Gamma r) r \, dr$ and $\int_0^a J_0(\Lambda r)I_0(\Gamma r)r\mathrm{d}r$ can be computed by \eqref{eq:bessel2} and \eqref{eq:bessel5} respectively.

\begin{theorem}\label{thm:kappavalue}
With notations defined in \eqref{eq:auxint1}, the entries of $\bm{\kappa}$ are expressed as follows
\begin{equation}\label{formulakappa}
\kappa_s=\left\{
\begin{aligned}
&\dfrac{2\pi k\Delta n}{n_0}\left(B^2\sum\limits_{l=-\infty}^{-1}\varsigma_{l,s}+AB\tau_s+B^2\sum\limits_{l=1}^{s-1}\zeta_{l,s}+B^2\sum\limits_{l=s+1}^{+\infty}\varsigma_{l,s}\right),\quad s>0,\\
&\dfrac{2\pi B^2k\Delta n}{n_0}\left(\sum\limits_{l=-\infty}^{-1}\varsigma_{l,0}+\sum\limits_{l=1}^{+\infty}\varsigma_{l,0}\right),\quad s=0.
\end{aligned}
\right.
\end{equation}
\end{theorem}

\begin{proof}
To calculate $\kappa_s$, we firstly consider the integral
\begin{equation}\label{Singleintegral}
I_{l,s}:=\iint_{B_a(lD_h,0)}\phi_0(x,y)\phi_s(x,y)\mathrm{d}x \mathrm{d}y
\end{equation}
for each $l\neq s$. In polar coordinates, $I_{l,s}$ is written as
$$
\begin{aligned}
I_{l,s}=B^2\int_0^{2 \pi }\mathrm{d}\theta\int_0^aK_0\left(\Gamma \sqrt{(lD_h+r\cos\theta)^2+(r\sin \theta)^2}\right)K_0\left(\Gamma\sqrt{[(l-s)D_h+r\cos\theta]^2+(r\sin\theta)^2}\right)r\mathrm{d}r
\end{aligned}
$$
for $l\neq 0$, and
$$
\begin{aligned}
I_{0,s}=AB\int_0^{2 \pi }\mathrm{d}\theta\int_0^aJ_0(\Lambda r)K_0(\Gamma\sqrt{(-sD_h+r\cos\theta)^2+(r\sin\theta)^2})r\mathrm{d}r
\end{aligned},
$$
for $l=0$.
\begin{itemize}
\item {\bf Case I: $l> 0$.} Applying the addition theorem \ref{Additiontheorem}, when $r<a$, it follows that
$$
\begin{aligned}
K_0(\Gamma \sqrt{(lD_h+r\cos\theta)^2+(r\sin \theta)^2})=\sum\limits_{m=-\infty}^{+\infty}(-1)^mI_m(\Gamma r)K_m(\Gamma lD_h)\cos(m\theta).
\end{aligned}
$$
Similarly,
$$
\begin{aligned}
K_0\left(\Gamma\sqrt{[(l-s)D_h+r\cos\theta]^2+(r\sin\theta)^2}\right)=\left\{
\begin{aligned}
&\sum\limits_{m=-\infty}^{+\infty}(-1)^mI_m(\Gamma r)K_m(\Gamma (l-s)D_h)\cos(m\theta),&& l>s\\
&\sum\limits_{m=-\infty}^{+\infty}I_m(\Gamma r)K_m(\Gamma (s-l)D_h)\cos(m\theta), && l<s.
\end{aligned}\right.
\end{aligned}
$$
Therefore, when $l>s$, the integral
$$
\begin{aligned}
I_{l,s}=B^2\int_0^{2 \pi }\mathrm{d}\theta\int_0^a\left(\sum\limits_{m=-\infty}^{+\infty}(-1)^mI_m(\Gamma r)K_m(\Gamma lD_h)\cos(m\theta) \right)\left(\sum\limits_{n=-\infty}^{+\infty}(-1)^nI_n(\Gamma r)K_n(\Gamma (l-s)D_h)\cos(n\theta)\right)r\mathrm{d}r.
\end{aligned}
$$
The integration over $\theta$ implies that only the terms with $\vert m\vert =\vert n\vert$ survive. Using the fact that $\forall m\in\mathbb{Z}, I_{-m}(z) = I_m(z)$ and $K_{-m}(z) = K_m(z)$, we have
$$
\begin{aligned}I_{l,s} = 2\pi B^2 \bigg[ K_0(\Gamma l D_h) K_0(\Gamma (l-s) D_h) \int_0^a I_0(\Gamma r)^2 r \, dr + 2\sum_{m=1}^\infty K_m(\Gamma l D_h) K_m(\Gamma (l-s) D_h) \int_0^a I_m(\Gamma r)^2 r \, dr \bigg].
\end{aligned}
$$
When $l<s$, the addition theorem \ref{Additiontheorem} gives the formula
$$
\begin{aligned}
I_{l,s}=B^2\int_0^{2 \pi }\mathrm{d}\theta\int_0^a\left(\sum\limits_{m=-\infty}^{+\infty}(-1)^mI_m(\Gamma r)K_m(\Gamma lD_h)\cos(m\theta) \right)\left(\sum\limits_{n=-\infty}^{+\infty}I_n(\Gamma r)K_n(\Gamma (s-l)D_h)\cos(n\theta)\right)r\mathrm{d}r.
\end{aligned}
$$
By integrating over $\theta$ first, we have
$$
\begin{aligned}
I_{l,s}=&2\pi B^2\bigg[K_0(\Gamma l D_h)K_0(\Gamma(s-l)D_h)\int_0^aI_0^2(\Gamma r)r \mathrm{d}r+2\sum\limits_{m=1}^\infty(-1)^mK_m(\Gamma l D_h)K_m(\Gamma(s-l)D_h)\int_0^aI_m^2(\Gamma r)r\mathrm{d}r\bigg].
\end{aligned}
$$
We can use the same steps to deal with the remaining cases.
\item {\bf Case II: $l<0$.}
Since $s\geqslant 0$, $l<s$, so the integral is
$$
\begin{aligned}
I_{l,s}=&2\pi B^2\bigg[K_0(\Gamma \vert l\vert D_h)K_0(\Gamma(l-s)D_h)\int_0^aI_0^2(\Gamma r)r \mathrm{d}r+2\sum\limits_{m=1}^\infty K_m(\Gamma \vert l\vert D_h)K_m(\Gamma(l-s)D_h)\int_0^aI_m^2(\Gamma r)r\mathrm{d}r\bigg].
\end{aligned}
$$
\item {\bf Case III: $l= 0$.}
The integral is
$$
I_{0,s}=2\pi ABK_0(\Gamma s D_h)\int_0^a J_0(\Lambda r)I_0(\Gamma r)r\mathrm{d}r,\quad s\neq 0.
$$
\end{itemize}
Finally, invoking the definition of $\kappa_{ij}$ in~\eqref{defkappa}, we arrive at the expression~\eqref{formulakappa}.
  \end{proof}


\section{Calculate the Band Structure}
In this section, we consider the NCME for an infinite waveguide system. Thus, the range of $m$ becomes $\mathbb{Z}$.
\subsection{Properties of the Matrices in NCME }
Before generalizing NCME to an infinite waveguide system, we firstly prove some useful lemmas.
\begin{proposition}\label{prop-4.1}For some fixed $(x,y)$, the sequence $\phi_m(x,y)$ decays exponentially as $m\to \infty$, namely, for $\vert m\vert$ large enough, there exist positive constants $C^\phi_1,C^\phi_2$ such that
 $$\phi_m(x,y)\leqslant C^\phi_1e^{-C^\phi_2\vert m\vert}.$$
 \end{proposition}
 \begin{proof}Recall that $\phi_m(x,y)=\phi(r_m)$ where $r_m=\sqrt{(x-mD_h)^2+y^2}$.
 Since $\lim\limits_{m\to \infty} r_m=+\infty$, one can assume that $r_m>a$ when $\vert m\vert$ greater than some $m_0$. Therefore, using the property of $K_0(z)$ \ref{lem-propK_0leq}, one obtains
 $$\phi(r_m)=BK_0(\Gamma r_m)\leqslant B \sqrt{\dfrac{\pi}{2\Gamma r_m}}e^{-\Gamma r_m}\leqslant B\sqrt{\dfrac{\pi}{2\Gamma a}}e^{-\Gamma r_m}=C^{\phi}_1e^{-\Gamma r_m}$$  where the constant $C^{\phi}_1=B\sqrt{\dfrac{\pi}{2\Gamma a}}$. As for $r_m$, $$r_m=\sqrt{(x-mD_h)^2+y^2}\geqslant \vert x-mD_h\vert= \vert m\vert D_h\left(1-\dfrac{\vert x\vert }{\vert m\vert D_h}\right).$$
Since $\lim\limits_{m\to\infty}\dfrac{\vert x\vert}{\vert m\vert D_h}$=0, there exists $m_1$ such that $\dfrac{\vert x\vert}{\vert m\vert D_h}\leqslant \dfrac{1}{2}$ for $\vert m\vert$  greater than $m_1$. Consequently, we have $r_m\geqslant \dfrac{D_h}{2}\vert m\vert $.
If we denote the constant $\Gamma \dfrac{D_h}{2}$ by $C^{\phi}_2$, then there holds $$\phi_m(x,y)\leqslant C^{\phi}_1e^{-C^{\phi}_2\vert m\vert}$$ for $\vert m\vert $ large enough as desired.
 \end{proof}

 \begin{theorem} \label{thm-estimateSxi}There exist two positive constants $C^S_1$, $C^S_2$ such that the following estimate holds for any non-negative integer $\xi$: $$S_\xi \leqslant C^S_1 e^{-C^S_2 \xi}$$
\end{theorem}
\begin{proof}By Proposition \ref{prop-3.1} and the proof of Proposition \ref{prop-4.1}, the following estimates for $\phi$ hold:
\begin{equation}\label{ineqrgeqa}\forall r\geqslant a,\quad\phi(r)\leqslant C_0e^{-\Gamma r},\end{equation}
\begin{equation}\label{ineqrleqa}\forall 0<r<a, \quad \phi(r)\leqslant A,\end{equation}
where $C_0=B\sqrt{\frac{\pi}{2\Gamma a}}$ is a positive constant.
The idea of estimating $$S_\xi =\iint_{\Omega}\phi_0(x,y)\phi_\xi (x,y)\mathrm{d}x\mathrm{d}y$$ is to split the integration domain into two regions. On each region one factor of the integrand is uniformly bounded, while the other provides exponential decay.

Define an auxiliary integral $$A_\gamma:=\iint_{\Omega}\phi(r)e^{\gamma r}\mathrm{d}x\mathrm{d}y,$$
where $r=\sqrt{x^2+y^2}$ and $0<\gamma<\Gamma$. $A_\gamma$ is well-defined, because
$$\begin{aligned}A_\gamma&=2\pi\int_0^\infty\phi(r)e^{\gamma r}r\mathrm{d}r= 2\pi\int_0^a\phi(r)e^{\gamma r}r\mathrm{d}r+2\pi\int_a^\infty\phi(r)e^{\gamma r}r\mathrm{d}r\\
&\leqslant 2\pi Ae^{\gamma a}\int_0^a r\mathrm{d}r+2\pi C_0\int_a^\infty e^{-(\Gamma-\gamma)r}r\mathrm{d}r=2\pi A\dfrac{a^2}{2}e^{\gamma a}+2\pi C_0\left(\dfrac{a}{\Gamma-\gamma}+\dfrac{1}{(\Gamma-\gamma)^2}\right)e^{-a(\Gamma-\gamma)}<+\infty.\end{aligned}$$
 Let $ u = x - \xi  D_h, v = y , r_\xi  := \sqrt{u^2 + v^2}, r_0 := \sqrt{(u + \xi  D_h)^2 + v^2}$, then $S_\xi$ is equivalent to
$$
\begin{aligned}
S_\xi  &= \iint_{\Omega} \phi(r_0) \phi(r_\xi ) \mathrm{d}u \mathrm{d}v&= \iint_{\{(u,v)\vert r_\xi \geqslant a\}} \phi(r_0) \phi(r_\xi ) \mathrm{d}u \mathrm{d}v+ \iint_{\{(u,v)\vert r_\xi \leqslant a\}} \phi(r_0) \phi(r_\xi ) \mathrm{d}u \mathrm{d}v =: I_1+I_2.
\end{aligned}
$$
We deal with the integral $I_1$ first.  By the triangle inequality, we have $r_0\geqslant \xi D_h-r_\xi$, therefore it follows that
 $$e^{-\gamma r_\xi }\leqslant e^{-\gamma(\xi D_h-r_0)}=e^{-\gamma \xi D_h}e^{\gamma r_0}.$$
 Since $r_\xi \geqslant a$, using \eqref{ineqrgeqa} the integrand can be estimated as
 $$\phi(r_0)\phi(r_\xi )\leqslant \phi(r_0)C_0e^{-\Gamma r_\xi }\leqslant \phi(r_0)C_0e^{-\gamma r_\xi }\leqslant \phi(r_0)C_0e^{-\gamma \xi D_h}e^{\gamma r_0}.$$
 Thus, $$\begin{aligned}
 I_1&=\iint_{\{(u,v)\vert r_\xi \geqslant a\}}\phi(r_0)\phi(r_\xi )\mathrm{d}u\mathrm{d}v\leqslant \iint_{\{(u,v)\vert r_\xi \geqslant a\}} \phi(r_0)C_0e^{-\gamma \xi D_h}e^{\gamma r_0}\mathrm{d}u\mathrm{d}v\\
 &\leqslant C_0e^{-\gamma \xi D_h} \iint_{\Omega} \phi(r_0)e^{\gamma r_0}\mathrm{d}u\mathrm{d}v
 =C_0e^{-\gamma \xi D_h} A_\gamma=C_{\gamma,1}e^{-C_{\gamma,2} \xi },\end{aligned}$$
 where $C_{\gamma,1}=C_0A_\gamma$ and $C_{\gamma,2}=\gamma D_h$ are constants related to $\gamma$ which can be any positive number less than $\Gamma$. Here we use the translation invariance of Lebesgue integral in the penultimate equality.

Next, we deal with the integral $I_2$. Since the region satisfies $r_\xi =\sqrt{u^2+v^2}\leqslant a$, it follows that $\vert u\vert+\vert v\vert\leqslant 2a$. Therefore,
$$r_0\geqslant \xi D_h-(\vert u\vert+\vert v\vert)\geqslant \xi D_h-2a.$$
When $\xi $ larger than $\dfrac{3a}{D_h}$, we have $r_0>a$, which means one can estimate $\phi(r_0)$ by \eqref{ineqrgeqa}
$$\phi(r_0)\leqslant C_0e^{-\Gamma r_0}\leqslant C_0e^{-\Gamma (\xi D_h-2a)}=C_0e^{-\Gamma \xi D_h}e^{\Gamma 2a}.$$
Thus, $$\begin{aligned}I_2&=\iint_{\{(u,v)\vert r_\xi \leqslant a\}} \phi(r_0) \phi(r_\xi ) \mathrm{d}u \mathrm{d}v\leqslant \iint_{\{(u,v)\vert r_\xi \leqslant a\}} C_0e^{-\Gamma \xi D_h}e^{\Gamma 2a}A\mathrm{d}u \mathrm{d}v=\pi a^2C_0Ae^{\Gamma 2a}e^{-\Gamma \xi D_h}\\
&\leqslant\pi a^2C_0Ae^{\Gamma 2a}e^{-\gamma \xi D_h}=C_{0,1}e^{-C_{\gamma, 2}\xi },  \end{aligned}$$
where $C_{0,1}=\pi a^2C_0Ae^{\Gamma 2a}$ and $C_{\gamma,2}=\gamma D_h$ are two constants.
For those $\xi\leqslant\dfrac{3a}{D_h}$, we consider the following maximum of a finite set
$$M:=\max_{\{\xi \in\mathbb{Z}_{\geqslant 0}\;\mid\; \xi \leqslant \frac{3a}{D_h}\}} \{ S_\xi  e^{C_{\gamma,2}\xi }\} <\infty.$$
Now we define constants $C^S_1:=\max\{ M,C_{0,1}+C_{\gamma,1}\}$ and $C^S_2:=C_{\gamma,2}$, and hence
$$S_\xi \leqslant C^S_1e^{-C^S_{2}\xi },\quad \forall \xi \in\mathbb{Z}_{\geqslant 0}.$$ The constants $C^S_1$, $C^S_2$ only depend on $\gamma$, where $0<\gamma<\Gamma$ is arbitrary
\end{proof}

\begin{theorem}\label{thm-4.3}
There exist two positive constants $C^{\kappa}_1$ and $C^{\kappa}_2$ such that for any  nonnegative integer $\xi=\vert i-j\vert$, the coupling coefficient
$\kappa_{\xi}$
satisfying the exponential decay estimation
$$\vert\kappa_{\xi}\vert\leqslant C^{\kappa}_1\, e^{-C^{\kappa}_2 \xi}.$$
\end{theorem}
\begin{proof} Since $$\kappa_\xi=\dfrac{k\Delta n}{n_0}\iint_{\bigcup\limits_{l\neq \xi}B_a(\xi D_h,0)}\phi_0(x,y)\phi_\xi(x,y)\mathrm{d}x\mathrm{d}y,$$
one can estimate $\kappa_\xi$ by using the estimate of $S_\xi$ in Theorem \ref{thm-estimateSxi}. Since the integration domain of $\kappa_\xi$ is a subset of $\Omega$,
$$\vert\kappa_\xi\vert
\leqslant \dfrac{k\Delta n}{n_0}\left\vert\iint_{\Omega}\phi_0(x,y)\phi_\xi(x,y)\mathrm{d}x\mathrm{d}y\right\vert=\dfrac{k \Delta n}{n_0}\vert S_\xi\vert
\leqslant \dfrac{k\Delta n}{n_0}C^S_1e^{-C^S_2\xi}.$$
Denote the constant $\dfrac{k\Delta n}{n_0}C^S_1$ and $C^S_2$ by $C^\kappa_1$ and $C^\kappa_2$ respectively. Consequently,$$
\vert \kappa_{\xi}\vert \leqslant C_1^{\kappa} e^{-C_2^{\kappa}\xi}.$$\end{proof}
From the relation $K_{nm}=\beta_0S_{nm}+\kappa_{nm}$ one can obtain the following corollary from above theorem.
\begin{corollary}There exist two positive constants $C^{K}_1$ and $C^{K}_2$ such that for any  nonnegative integer $\xi=\vert i-j\vert$, the coupling coefficient
$K_{\xi}$
satisfying the exponential decay estimate
$$\vert K_{\xi}\vert\leqslant C^{K}_1\, e^{-C^{K}_2 \xi}.$$
\end{corollary}
\begin{remark}In fact, these three constants $C^{S}_2, C^{\kappa}_2$ and $C^{K}_2$ are the same. Thus, we denote them by $C_2$ for convenience.
\end{remark}

\subsection{NCME for Infinite Waveguide System}
For an infinite waveguide system, the coupled mode equation should be composed by  operators in some Hilbert space. Starting from the finite-dimensional coupled-mode system, we naturally extend it to $\ell^2(\mathbb{Z};\mathbb{C})$ by defining the operators $\mathcal{S}, \mathcal{K}$ as follows.

 Let $\mathcal{H}:=\ell^2(\mathbb{Z};\mathbb{C})$ equipped with inner product $\langle \bm{c}, \bm{d} \rangle = \sum_{m \in \mathbb{Z}} c_m\overline{d_m}$
 for two sequences $\bm{c}:=\{c_m\}_{m \in \mathbb{Z}}$ and $\bm{d}:=\{d_m\}_{m \in \mathbb{Z}}$ in $\mathcal{H}$.

Define two operators $\mathcal{S}, \mathcal{K}:\mathcal{H}\rightarrow \mathcal{H}$
by \begin{equation}(\mathcal{S}\bm{c})_n=\sum\limits_{m\in\mathbb{Z}}S_{\vert n-m\vert} c_m=\sum\limits_{\xi '\in\mathbb{Z}}S_{\vert \xi'\vert} c_{n-\xi'}\end{equation}
and \begin{equation}(\mathcal{K}\bm{c})_n=\sum\limits_{m\in\mathbb{Z}}K_{\vert n-m\vert} c_m=\sum\limits_{\xi '\in\mathbb{Z}}K_{\vert \xi'\vert} c_{n-\xi'}\end{equation}
respectively with $\xi'=n-m$. It is immediate that $\mathcal{S}$ and $\mathcal{K}$ are self-adjoint by B{\"o}ttcher~\cite[p.~3, Proposition~1.3]{bottcher2000toeplitz}.

\begin{remark}When we consider the finite waveguide system, the formulas $(\mathcal{S}\bm{c})_n=\sum\limits_{m\in\mathbb{Z}}S_{\vert n-m\vert} c_m$ and $(\mathcal{K}\bm{c})_n=\sum\limits_{m\in\mathbb{Z}}K_{\vert n-m\vert} c_m$ can be more easily reduced to the finite case rather than use   $(\mathcal{S}\bm{c})_n=\sum\limits_{\xi '\in\mathbb{Z}}S_{\vert \xi'\vert} c_{n-\xi'}$
and $(\mathcal{K}\bm{c})_n=\sum\limits_{\xi '\in\mathbb{Z}}K_{\vert \xi'\vert} c_{n-\xi'}$.
\end{remark}
We first check the well-definedness of $\mathcal{S}$ and $\mathcal{K}$ in the following proposition.
\begin{proposition}\label{prop:skwelldefinedness}The operators $\mathcal{S}$ and $\mathcal{K}$ are well-defined, namely,
$(\mathcal{S}\bm{c})_n=\sum\limits_{\xi '\in\mathbb{Z}}S_{\vert \xi'\vert} c_{n-\xi'}$
and $(\mathcal{K}\bm{c})_n=\sum\limits_{\xi '\in\mathbb{Z}}K_{\vert \xi'\vert} c_{n-\xi'}$ are convergent and the sequences $\{(\mathcal{S}\bm{c})_n\}_{n\in \mathbb{Z}}$ and $\{(\mathcal{K}\bm{c})_n\}_{n\in\mathbb{Z}}$ belong to $\mathcal{H}$.
\end{proposition}
\begin{proof}
By theorem \ref{thm-estimateSxi} and Cauchy-Schwarz inequality,
$$
\begin{aligned}
\vert (\mathcal{S}\bm{c})_n\vert &\leqslant \sum\limits_{\xi'\in\mathbb{Z}}S_\xi\vert c_{n-\xi'}\vert
\leqslant \sum\limits_{\xi'\in\mathbb{Z}}C_1^Se^{-C_2\xi}\vert c_{n-\xi'}\vert \\
&\leqslant C_1^S\left(\sum\limits_{\xi'\in\mathbb{Z}}e^{-2C_2\xi}\right)^{1/2} \left(\sum\limits_{\xi'\in\mathbb{Z}}\vert c_{n-\xi'}\vert^2\right)^{1/2}=C_1^S\left(\dfrac{1+e^{-2C_2}}{1-e^{-2C_2}}\right)^{1/2}\Vert\bm{c}\Vert_{\mathcal{H}},
\end{aligned}
$$
so the series for $(\mathcal{S}\bm{c})_n$ converges absolutely for each $n\in\mathbb{Z}$.

Similarly, for $\mathcal{K}$ we have
$$\vert (\mathcal{K}\bm{c})_n\vert\leqslant C_1^K\left(\dfrac{1+e^{-2C_2}}{1-e^{-2C_2}}\right)^{1/2}\Vert\bm{c}\Vert_{\mathcal{H}}. $$
Thus, the series for $(\mathcal{K}\bm{c})_n$ also converges absolutely for each $n\in\mathbb{Z}$.

To show that $\{(\mathcal{S}\bm{c})_n\}_{n \in \mathbb{Z}} \in \mathcal{H}$, we need
$$\Vert\mathcal{S}\bm{c}\Vert_\mathcal{H}^2 = \sum_{n \in \mathbb{Z}} \vert(\mathcal{S}\bm{c})_n\vert^2 < \infty.$$
Define a new sequence $S'=\{S'_{\xi'}\}_{\xi'\in\mathbb{Z}}$ by $S'_{\xi'}:=S_\xi$ with $\xi=\vert\xi'\vert$, which belongs to $\ell^1(\mathbb{Z})$ since
$$\Vert S'\Vert_{\ell^1} = \sum_{\xi' \in \mathbb{Z}} S_{|\xi'|} \leqslant  C^S_1 \left( 1 + \frac{2 e^{-C_2}}{1 - e^{-C_2}} \right).$$ Then $(\mathcal{S}\bm{c})_n$ can be regarded as the convolution of $S'$ and $\bm{c}$ at $n$:
$$(\mathcal{S}\bm{c})_n=\sum\limits_{\xi'\in\mathbb{Z}}S'_{\xi'}c_{n-\xi'}= ( S'\ast \bm{c})_n.$$
By Young’s inequality for convolutions \cite[Eq.~1.1]{charalambides2011near},
$$\Vert\mathcal{S}\bm{c}\Vert_\mathcal{H} = \Vert S' \ast \bm{c}\Vert_{\ell^2} \leq \Vert S'\Vert_{\ell^1} \Vert \bm{c}\Vert_{\ell^2} \leq C^S_1 \left( 1 + \frac{2 e^{-C_2}}{1 - e^{-C_2}} \right) \Vert\bm{c}\Vert_\mathcal{H} < \infty.$$
Hence, $\mathcal{S}\bm{c} \in \mathcal{H}$. For $\mathcal{K}$, we define $\mathcal{K}'$ similarly and conclude that $\mathcal{K}\bm{c} \in\mathcal{H}$ with similar reasoning.
\end{proof}

From the proof above, we also obtain two corollaries:
\begin{corollary}\label{cor-SKbdd}The operators $\mathcal{S}$ and $\mathcal{K}$ are bounded on $\mathcal{H}$.
\end{corollary}
\begin{corollary}The sequences $\{ S'_{\xi'}\}_{\xi'\in\mathbb{Z}}$ and $\{K'_{\xi'}\}_{\xi'\in\mathbb{Z}}$ are $\ell^1(\mathbb{Z})$.
\end{corollary}

Similar to \eqref{eq:totalenvelope}, we assume that the amplitude function of the infinite waveguide array is a superposition of these modes with $z$-dependent amplitudes $$\psi(x,y,z)=\sum\limits_{m\in\mathbb{Z}}c_m(z)\phi_m(x,y),$$
where $\phi_m(x,y)=\phi_0(x-mD_h,y)$ and the mode coefficient $\{c_m(z)\}$ is regarded as an infinite vector $\bm{C}(z)$ in $\mathcal{H}$.

\begin{proposition}For any fixed point $(x,y,z)$, the series  $$\psi(x,y,z)=\sum\limits_{m=-\infty}^{+\infty}c_m(z)\phi_m(x,y)$$ converges absolutely.
\end{proposition}
\begin{proof}Choose a sufficiently large $m_0$ such that $\phi_m(x,y)$ decays exponentially for $\vert m\vert>m_0$.
The series can be decomposed into two parts:
$$\sum\limits_{\vert m\vert\leqslant m_0}c_m(z)\phi_m(x,y)+\sum\limits_{\vert m\vert>m_0}c_m(z)\phi_m(x,y).$$
The first part must be finite, so we focus on the second part. Cauchy-Schwarz inequality implies
$$\sum_{|m|>m_0} |c_m(z)| |\phi_m(x,y)| \le \left(\sum_{|m|>m_0} |c_m(z)|^2\right)^{1/2} \left(\sum_{|m|>m_0} |\phi_m(x,y)|^2\right)^{1/2}.$$ The first factor is convergent because $\{c_m(z)\}$ is in $\mathcal{H}$ for any fixed $z$, and for the second factor we have
$$\sum_{|m|>m_0} |\phi_m(x,y)|^2\leqslant \sum_{|m|>m_0}C_1^2e^{-2C_2\vert m\vert}<+\infty,$$
by the proposition \ref{prop-4.1}. Hence the series defining $\psi(x,y,z)$ is absolutely convergent.
\end{proof}
The following proposition ensures that the energy of the wave function is finite.
\begin{proposition}The function $\psi(x,y,z)$ belongs to $L^2(\Omega)$ for any $z\in\mathbb{R}_{\geqslant 0}$.
\end{proposition}
\begin{proof}
For fixed $z$,
$$\vert\psi(x, y, z)\vert^2 =\sum_{m, n\in\mathbb{Z} }c_m(z)\overline{c_n(z)} \phi_m(x, y) \phi_n(x, y).$$

Integrating over $\Omega=\mathbb{R}^2$ yields
$$
\begin{aligned}\Vert\psi(\cdot, \cdot, z)\Vert_{L^2}^2 &= \sum_{m, n=-\infty}^{+\infty} c_m(z) \overline{c_n(z)} \int_{\Omega} \phi_m(x, y) \phi_n(x, y) \, \mathrm{d}x \mathrm{d}y = \sum_{m, n=-\infty}^{+\infty} c_m(z) \overline{c_n(z)} S_{mn}=\langle\bm{c},\mathcal{S}\bm{c}\rangle.
\end{aligned}
$$
The boundedness of $\mathcal{S}$ immediately implies that $\psi\in L^2(\mathbb{R}^2)$.
\end{proof}

We denote by $C_F^1([0,+\infty), \mathcal{H})$ the space of all maps $[0,+\infty)\xlongrightarrow{\bm{C}}\mathcal{H}$ that are Fr\'echet differentiable on $[0,+\infty)$ in the sense above, with continuous derivative.
\begin{remark}There exists an isometry $L(\mathbb{R}, \mathcal{H})\to\mathcal{H}$ given by $(\mathbb{R}\xlongrightarrow{T} H)\mapsto T(1)$, then $L(\mathbb{R}, \mathcal{H})\cong \mathcal{H}$. Consequently, the Fr\'{e}chet derivative $D_F \bm{C}(z)$ of the map $\bm{C}: [0, \infty) \to \mathcal{H}$ at $z$ can be identified with an element in $\mathcal{H}$.
\end{remark}
For any $\bm{C}\in C_F^1([0,+\infty), \mathcal{H}) $, similar to the derivation in the finite-dimensional setting (where the $\psi=\sum c_m(z)\phi_m$ is substituted into the paraxial wave equation), one arrives at the following operator form for the infinite waveguide system:
\begin{equation}\label{eq-GNCME}\ii\mathcal{S}D_F\bm{C}(z)+\mathcal{K}\bm{C}(z)=\bm{0}.
\end{equation}

\begin{theorem}\label{thm-Sinverti}The bounded linear operator $\mathcal{S}$ is invertible.
\end{theorem}
\begin{proof}Denote $\mathcal{I}:\mathcal{H}\rightarrow\mathcal{H}$ as the identity operator. According to the definition of $S_\xi$, $S_0 = 1$, so $\mathcal{S}=\mathcal{I}+\mathcal{R}$, where $\mathcal{R}:\mathcal{H}\rightarrow\mathcal{H}$ is defined by
$$(\mathcal{R}\bm{c})_n=\sum\limits_{\xi'\in\mathbb{Z}\setminus\{0\}}S'_{\xi'}c_{n-\xi'}.$$
For convenience, define a new sequence $S''=\{S''_{\xi'}\}_{\xi'\in\mathbb{Z}}$ by
$$S''_{\xi'}=\left\{\begin{aligned}S'_{\xi'}&,\quad \xi'\neq 0\\0&,\quad\xi'=0\end{aligned}\right.$$
Then $\Vert S''\Vert_{\ell^1(\mathbb{Z})}=\sum\limits_{\xi'\in\mathbb{Z}}\vert S''_{\xi'}\vert$.
By the Young's inequality,
$$\Vert \mathcal{R}\bm{c}\Vert_\mathcal{H}
=\Vert S''\ast\bm{c}\Vert_{\ell^2(\mathbb{Z})}
\leqslant \Vert S''\Vert_{\ell^1(\mathbb{Z})}\Vert \bm{c}\Vert_{\mathcal{H}}.$$ Therefore, the norm of the operator $\mathcal{R}$ can be bounded as
$$\Vert \mathcal{R}\Vert\leqslant \sum\limits_{\xi'\in\mathbb{Z}}S''_{\xi'}=:\eta.$$
By the Neumann series theorem, to prove that $\mathcal{S}$ is invertible, one only needs to show $\Vert\mathcal{R}\Vert<1$, which is implied by $\eta<1$. Hence, we want to find a bound for $\eta$. Since the constant $C_1^S$ in the theorem \ref{thm-estimateSxi} is so large that we cannot reach  $\eta<1$ directly, we estimate $\eta$ using a trick in the following:
$$\begin{aligned}
\eta&=2\sum\limits_{\xi=1}^{\infty}S_\xi=2\left[S_1+\cdots+S_{10}+\sum\limits_{\xi=11}^{\infty}S_\xi\right]\\
&\leqslant 2\left[S_1+\cdots+S_{10}+\sum\limits_{\xi=11}^{\infty}C_1^Se^{-C^S_2\xi}\right]= 2\left[S_1+\cdots+S_{10}+C^S_1\dfrac{e^{-11\cdot C^S_2}}{1-e^{-C^S_2}}\right].
\end{aligned}$$
Since $\gamma$ can be chosen arbitrarily in Theorem~\ref{thm-estimateSxi}, we may fix $\gamma$ such that $\eta<1$. 
Hence the operator $\mathcal{S}=I+\mathcal{R}$ satisfies 
$\Vert \mathcal{R}\Vert<1$ and is therefore invertible.\end{proof}
\begin{remark}
As an example to evaluate this bound of $\eta$, we choose $\gamma=10^5$ in the theorem \ref{thm-estimateSxi} and calculate the numerical integral of $A_\gamma$, and the formula \eqref{formulaSh} is applied to compute $S_\xi$, $\xi=1,\cdots,10$. With parameters set as Remark \ref{rmk:1} and Remark \ref{rmk:2}, the numerical calculation shows that upper bound of $\eta$ is approximately $0.3951$, which is smaller than $1$.
\end{remark}

Thus, the operator form equation can be expressed as
\begin{equation}\label{eq-GNCME2}\ii D_F\bm{C}(z)+\mathcal{W}\bm{C}(z)=\bm{0},
\end{equation}
where  $\mathcal{W}=\mathcal{S}^{-1}\mathcal{K}$.  By the corollary \ref{cor-SKbdd}, we know that $\mathcal{W}$ is bounded.
\begin{theorem}The operator $\mathcal{W}$ is self-adjoint. Consequently, the eigenvalues of the operator $\mathcal{W}$ are real-valued.
\end{theorem}
\begin{proof}Since the operators $\mathcal{S}$ and $\mathcal{K}$ are convolution operators, it follows that they are commutative: $\mathcal{SK}=\mathcal{KS}$, and hence $\mathcal{K}\mathcal{S}^{-1}=\mathcal{S}^{-1}\mathcal{K}$. Note that $\mathcal{S}$, $\mathcal{K}$ and $\mathcal{S}^{-1}$ are all self-adjoint, then
$$\langle \mathcal{S}^{-1}\mathcal{K}\bm{x},\bm{y}\rangle=\langle \mathcal{K} \mathcal{S}^{-1} \bm{x}, \bm{y} \rangle=\langle \mathcal{S}^{-1}\bm{x},\mathcal{K}\bm{y}\rangle=\langle \bm{x},\mathcal{S}^{-1}\mathcal{K}\bm{y}\rangle$$
\end{proof}

\subsection{Definition of the dispersion relation }
\begin{definition}\label{def-continuum}
The {\bf continuum} of the equation \eqref{eq-GNCME2} is defined as the  spectrum of $\mathcal{W}$, denoted by $\mathcal{C}(\mathcal{W})$.
\end{definition}
By Lemma~\ref{lem-DAVIES} given by \cite[p.~65, Theorem.~2.4.4]{davies2007linear}, if we verify that $\mathcal{W}$ is a convolution operator with kernel $W\in\ell^1(\mathbb{Z})$, then we only need to calculate its  discrete  Fourier transformation of $\mathcal{W}$ for seeking  the continuum $\mathcal{C}(\mathcal{W})$.
\begin{lemma}\label{lem-DAVIES} Let $\mathcal{A}:\ell^p(\mathbb{Z})\to \ell^p(\mathbb{Z})$ be the bounded operator $\mathcal{A}\bm{c}=A\ast \bm{c}$ where $A\in\ell^1(\mathbb{Z})$ and $1\leqslant p\leqslant \infty$. Then the spectrum of $\mathcal{A}$ coincides with the range of $\hat{a}(\theta)$, i.e. $$\operatorname{Spec}(\mathcal{A})=\{\hat{a}(\theta)\mid \theta\in \mathbb{T}\}.$$
\end{lemma}

\begin{definition}The {\bf dispersion relation} is the function
$$ \widehat{W}(\theta)=\sum\limits_{n\in\mathbb{Z}}W_ne^{-in\theta}, \qquad \theta\in\mathbb{T}.$$
\end{definition}
\begin{remark}The unit circle $\mathbb{T} = \{z\in\mathbb{C}:|z|=1\}$ is a compact abelian group under multiplication. It is canonically isomorphic to the additive group $\mathbb{R}/2\pi\mathbb{Z}$. Thus, one often parametrizes $\mathbb{T}$ by the interval $[-\pi,\pi]$ with $-\pi\sim\pi$.
\end{remark}
From \cite[p.~2, Example~3]{folland2016course},
$\ell^1(\mathbb{Z})$, equipped with pointwise addition, convolution as multiplication,
and the norm $\|\cdot\|_{\ell^1(\mathbb{Z})}$, is a unital commutative Banach algebra,
which we denote by $\mathscr{A}:=(\ell^1(\mathbb{Z}),\ast,\|\cdot\|_{\ell^1(\mathbb{Z})}).$
Moreover, the space of all bounded linear operators on $\ell^2(\mathbb{Z})$,
endowed with the usual operations and the operator norm,
is a unital Banach algebra, denoted by
$\mathscr{B}(\ell^2(\mathbb{Z})) = \mathscr{B}(\mathcal{H})$.
Define a map $\Phi: \mathscr{A}\rightarrow \mathscr{B}(\mathcal{H})$ by mapping $k\in\mathscr{A}$ to $\Phi(k)$ which satisfies
$$\forall\bm{c}\in\mathcal{H},\quad (\Phi(k)(\bm{c}))_n=(k\ast \bm{c})_n=\sum\limits_{m\in\mathbb{Z}}k_{n-m}c_m.$$
By Young's inequality, $\Phi(k)(\bm{c})\in \mathcal{H}$, and hence the map $\Phi$ is well-defined. Denote the range of $\Phi(\mathscr{A})$ by $\mathscr{C}$, which is the set of all convolution operators with kernels in $\ell^1(\mathbb{Z})$. The map $\Phi: \mathscr{A} \to \mathscr{B}(\mathcal{H})$ is a unital Banach algebra homomorphism, as it preserves addition, scalar multiplication, and multiplication, maps the unit $\delta_0$ to the identity operator $I$, and is contractive by Young's inequality $\Vert\Phi(k)\Vert_{\mathscr{B}(\mathcal{H})} \leq \|k\|_{\ell^1(\mathbb{Z})}$. Note that $(\Phi(k)(\delta_0))_n =k_n$, so $\Phi$ is injective. To conclude, we have
\begin{proposition}The map $\Phi$ is an injective algebra homomorphism, its image $\mathscr{C}$ is a subalgebra of $\mathscr{B}(\mathcal{H})$, and hence $\mathscr{A} \cong \mathscr{C}$ as algebras.
\end{proposition}
\begin{corollary}\label{cor-4.15}$k$ is invertible in $\mathscr{A}$ if and only if $\Phi(k)$ is invertible in $\mathscr{C}$.
\end{corollary}
Denote $\sigma(\mathscr{A})$ as the set of all multiplicative functionals on $\mathscr{A}$, which are nonzero homomorphisms from $\mathscr{A}$ to $\mathbb{C}$. The Gelfand transform maps $k\in\mathscr{A}$ to $\widehat{k}\in C(\sigma(\mathscr{A}))$ defined by $\widehat{k}(h)=h(k)$ for $h\in \sigma(\mathscr{A})$( see Folland~\cite[p.~7]{folland2016course}).
Notice that the Gelfand transform is a homomorphism from $\mathscr{A}$
to $C(\sigma(\mathscr{A}))$, with two important results (see Folland~\cite[p.~9, Theorem~1.17]{folland2016course}
and~\cite[p.~7, Theorem~1.13(b)]{folland2016course} or Katznelson~\cite[p.~217]{katznelson2004introduction}):
\begin{lemma}\label{lem-4.16}
$k$ is invertible if and only if $\hat{k}$ never vanishes.
\end{lemma}
\begin{lemma}\label{lem-4.17}
$\sigma(\mathscr{A})=\sigma(\ell^1(\mathbb{Z}))$ can be identified with the unit circle $\mathbb{T}$
in such a way that the Gelfand transform on $\mathscr{A}$ becomes Fourier transform:
$$\hat{k}(e^{-i\theta})=\sum\limits_{n\in\mathbb{Z}}k_ne^{-in\theta}.$$
\end{lemma}

We also write $\hat{k}(e^{-i\theta})$ as $\hat{k}(\theta)$ for $\theta \in [-\pi, \pi]$, which is homeomorphic to the unit circle $\mathbb{T}$. The function $\hat{k}(\theta)$ is called the Fourier symbol of $k$. By combining Corollary \ref{cor-4.15}, Lemma \ref{lem-4.16}, and Lemma \ref{lem-4.17}, we have the following theorem.

\begin{theorem}[Wiener's lemma]\label{thm-4.18}
The following are equivalent:
\begin{enumerate}
\item $\Phi(k)$ is invertible in $\mathscr{C}$;
\item $k$ is invertible in the Banach algebra $\mathscr{A}$;
\item $\hat{k}(\theta)\neq 0$ for all $\theta\in[-\pi,\pi)\cong\mathbb{T}$.
\end{enumerate}
\end{theorem}

\begin{proposition}
The operator $\mathcal{W}$ is a convolution operator with kernel $W\in\ell^1(\mathbb{Z})$.
\end{proposition}
\begin{proof}Since $\mathcal{W}=\mathcal{S}^{-1}\mathcal{K}$ and the operator $\mathcal{S}$ and $\mathcal{K}$ are convolution operator on $\mathcal{H}$ with kernel in $\ell^1(\mathbb{Z})$, which are $S'$ and $K'$ defined in the proof of Proposition \ref{prop:skwelldefinedness}, it follows that $\mathcal{S}=\Phi(S')$ and $\mathcal{K}=\Phi(K')$. By Theorem \ref{thm-4.18}, the invertibility of $\Phi(S')$ in $\mathscr{C}$ (Theorem~\ref{thm-Sinverti}) implies that there exists $U\in \mathscr{A}$ such that $S'\ast U=U\ast S'=\delta_0.$ Thus, $\Phi(S')^{-1}=\Phi(U)$. Using the algebra homomorphism property:
$$\mathcal{W}=\Phi(S')^{-1}\Phi(K')=\Phi(U\ast K').$$
Define $W=U\ast K'\in\mathscr{A}$, then $\mathcal{W}=\Phi(W)$. Therefore, the operator $\mathcal{W}$ is a convolution operator with kernel $W\in\ell^1(\mathbb{Z})$.
\end{proof}

Since the Gelfand transform is a homomorphism, we have
$$\widehat{U\ast S'}(\theta)=\widehat{U}(\theta)\widehat{S'}(\theta)=\widehat{\delta_0}(\theta)=1.$$ By the theorem \ref{thm-4.18}, $\widehat{S'}\neq 0$,
then $$\widehat{W}(\theta)=\widehat{U}(\theta)\widehat{K'}(\theta)=\dfrac{\widehat{K'}(\theta)}{\widehat{S'}(\theta)}.$$
Thus, we conclude that the dispersion relation is the function
 \begin{equation}\label{eq-4.7}
 \widehat{W}(\theta)
 =\dfrac{\widehat{K'}(\theta)}{\widehat{S'}(\theta)}
 =\dfrac{K_0+2\sum\limits_{m=1}^\infty K_m\cos(m\theta)}{S_0+2\sum\limits_{m=1}^\infty S_m\cos(m\theta)},\quad\theta\in [-\pi,\pi].\end{equation}

\subsection{The range of the dispersion relation}
Now we aim at showing the monotonicity of dispersion function $\widehat{W}(\theta)$ in first Brillouin zone $[0,\pi]$. We prove a few lemmas first.

\begin{lemma}\label{lem-4.20}Let $\{a_n\}_{n=0}^\infty$ and $\{b_n\}_{n=0}^\infty$ be two positive sequences which decay exponentially as $n$ increases. 
Let $c_n=n(b_na_0-b_0a_n)$ for $n\geqslant 1$. Assume that $c_1<0$, and $\vert c_1\vert >\sum\limits_{n=2}^\infty n\vert c_n\vert$. Define $M(\theta)=\sum\limits_{n=1}^\infty c_n\sin(n\theta),\theta\in [0,\pi]$. Then $M(\theta)<0$ for $\theta\in (0,\pi)$.
\end{lemma}

\begin{proof}Since the sequence $\{a_n\}_{n=0}^\infty$ and $\{b_n\}_{n=0}^\infty$  decay exponentially, the series $\sum\limits_{n=2}^\infty n\vert c_n\vert$ converges. By induction, one can show that
\begin{equation}\label{eq-4.8}\vert \sin(n\theta)\vert\leqslant n\vert \sin(\theta)\vert, \quad n\geqslant 1.\end{equation}
Thus, $$M(\theta)=c_1\sin\theta+\sum\limits_{n=2}^\infty c_n\sin(n\theta)\leqslant \sin\theta\left(-\vert c_1\vert+\sum\limits_{n=2}^\infty n\vert c_n\vert\right)<0.$$
\end{proof}
We define $$H(\theta):=S_a(\theta)C_b(\theta)-C_a(\theta)S_b(\theta),$$
where $S_a(\theta):=\sum\limits_{n=1}^\infty na_n\sin(n\theta)$, $S_b(\theta):=\sum\limits_{n=1}^\infty nb_n\sin(n\theta)$,  $C_a(\theta):=\sum\limits_{m=1}^\infty a_m\cos(m\theta)$ and $C_b(\theta):=\sum\limits_{m=1}^\infty b_m\cos(m\theta)$.

\begin{lemma}\label{lem-4.21}Use the same assumption as Lemma \ref{lem-4.20}. Let
$$
\Xi = \sum_{n=1}^\infty \sum_{m=1}^\infty a_n b_m |n^2 - m^2|= \sum_{n > m \geq 1} (n^2 - m^2) (a_n b_m + a_m b_n).
$$
If $\delta_c:=\vert c_1\vert-\sum\limits_{n=2}^\infty n\vert c_n\vert>2\Xi$, then $2M(\theta)+4H(\theta)<0$.
\end{lemma}
\begin{proof}Using the equation \eqref{eq-4.8}, we have
$$\vert M(\theta)\vert \geqslant \vert c_1\vert \sin\theta-\sum\limits_{n\geqslant 2}\vert c_n \vert\vert \sin(n\theta)\vert \geqslant \delta_c\sin\theta.$$
Note that
$$
\begin{aligned}
H(\theta)
&=\sum_{n,m\geqslant 1} a_n b_m\big( n\sin(n\theta)\cos(m\theta)-m\cos(n\theta)\sin(m\theta)\big) \\
&=\dfrac{1}{2}\sum_{n,m\geqslant1} a_n b_m\Big[(n-m)\sin((n+m)\theta)+(n+m)\sin((n-m)\theta)\Big].
\end{aligned}
$$
By \eqref{eq-4.8} again, we have
$$
\begin{aligned}
\vert H(\theta)\vert &=\dfrac{1}{2}\sum_{n,m\geqslant1} a_n b_m\Big[\vert n-m \vert\vert \sin((n+m)\theta)\vert +(n+m)\vert \sin((n-m)\theta)\vert \Big]\\
&\leqslant \sum_{n,m\geqslant1} a_n b_m\vert n^2-m^2\vert \sin(\theta)=\Xi \sin(\theta).
\end{aligned}
$$
Thus,
$$-M(\theta)=\vert M(\theta)\vert\geqslant \delta_c\sin(\theta)>2\Xi \sin(\theta)\geqslant 2\vert H(\theta)\vert,$$
which yields the desired result.
\end{proof}

 Define an operator on $\mathcal{H}$ denoted $\boldsymbol{\kappa}$ by $(\boldsymbol{\kappa}\bm{c})_n=\sum\limits_{m\in\mathbb{Z}}\kappa_{\vert n-m\vert} c_m=\sum\limits_{\xi '\in\mathbb{Z}}\kappa_{\vert \xi'\vert} c_{n-\xi'}$ for $n\geqslant 0$. We impose the following assumption on our system.

 \begin{assumption}\label{assumption1}Let $c_{S,\kappa}(n):=n(S_n\kappa_0-S_0\kappa_n)$, $n\geqslant 1$, $\Xi_{S,\kappa} := \sum_{n,m=1}^\infty \kappa_n S_m |n^2 - m^2|$. We require $\vert c_{S,\kappa}(1)\vert-\sum\limits_{n=2}^\infty n\vert c_{S,\kappa}(n)\vert>2\Xi_{S,\kappa}>0$ and $c_{S,\kappa}(1)<0$.
 \end{assumption}

\begin{remark}[Numerical verification of Assumption~\ref{assumption1}]
We numerically verify the inequalities in Assumption~\ref{assumption1} with parameters from Remark \ref{rmk:3}. The sequences $\{\kappa_n\}_{n\ge0}$ and
$\{S_n\}_{n\ge0}$ are computed from their defining integral representations
involving Bessel and modified Bessel functions, evaluated using
arbitrary-precision arithmetic with 20-digit precision. All infinite sums are
truncated at $N=10$, which is sufficient due to the exponential decay of the
coefficients.

The numerical calculation shows that
$$
-93.2238\approx c_{S,\kappa}(1)<0, \qquad
|c_{S,\kappa}(1)|>\sum_{n=2}^{N} n|c_{S,\kappa}(n)|\approx 37.6390,
$$
and moreover,
$$
55.5849\approx |c_{S,\kappa}(1)|-\sum_{n=2}^{N} n|c_{S,\kappa}(n)|>2\Xi_{S,\kappa}\approx 22.3723,
$$
which is consistent with the statement of Assumption~\ref{assumption1}.
\end{remark}

\begin{theorem}The dispersion relation $\widehat{W}(\theta)$ is an even function and monotonically decreasing on $[0,\pi]$.
\end{theorem}
\begin{proof} The evenness of the dispersion relation $\widehat{W}(\theta)$ follows from its expression in \eqref{eq-4.7}.  By the equation \eqref{eq:defskappa}, one can find that $$\widehat{W}(\theta)=\dfrac{\beta_0\widehat{S'}(\theta)+\widehat{\kappa'}(\theta)}{\widehat{S'}(\theta)}=\beta_0+\dfrac{\widehat{\kappa'}(\theta)}{\widehat{S'}(\theta)}.$$ 
Since the sequences $\{S_n\}_{n=0}^\infty$ and $\{\kappa_n\}_{n=0}^\infty$ decay exponentially as $n$ increases (by Theorems \ref{thm-estimateSxi} and \ref{thm-4.3}), all the associated series are differentiable.  It suffices to show that the derivative $\left(\frac{\widehat{\kappa'}(\theta)}{\widehat{S'}(\theta)}\right)' = \frac{D(\theta)}{(\widehat{S'}(\theta))^2}$ is negative on $(0, \pi)$, where $D(\theta)$ is given by
$$2\sum\limits_{n=1}^\infty n(S_n\kappa_0-\kappa_nS_0)\sin(n\theta)+4\sum\limits_{n=1,m=1}^\infty n(\kappa_nS_m-\kappa_m S_n)\sin(n\theta)\cos(m\theta).$$
Set $M_{S,\kappa}(\theta):=\sum\limits_{n=1}^\infty n(S_n\kappa_0-\kappa_nS_0)\sin(n\theta)$, $H_{S,\kappa}(\theta):=\sum\limits_{n=1,m=1}^\infty n(\kappa_nS_m-\kappa_m S_n)\sin(n\theta)\cos(m\theta).$ By Lemma \ref{lem-4.20}, \ref{lem-4.21} and Assumption \ref{assumption1}, we conclude that $$D(\theta)=2M_{S,\kappa}(\theta)+4H_{S,\kappa}(\theta)<0,\quad \forall\theta\in(0,\pi).$$ Therefore, the dispersion relation $\widehat{W}(\theta)$ is monotonically decreasing on $[0,\pi]$.
\end{proof}
From the theorem above, we know that the maximum and minimum of the dispersion relation $\widehat{W}(\theta)$ on $[-\pi,\pi]$ are $\widehat{W}(0)$ and $\widehat{W}(\pi)$ respectively.
The values of the horizontal dispersion function $\widehat{W}(\theta)$ at $\theta = 0$ and $\theta = \pi$ were computed numerically using a truncated Fourier-type summation over the coupling coefficients $K_m$ and normalization factors $S_m$.
\begin{remark}\label{rmk:9}
With the physical parameter in Remark \ref{rmk:3}, the continuum band of the horizontal waveguide system(that we are interested in) is $\mathcal{C}(\mathcal{W})=[560.035822,962.112305]$.
\end{remark}

\section{Occurrence of BICs}

From a physical point of view, any wave whose propagation constant lies in the continuous spectrum should propagate to infinity in a waveguide system. However, by carefully designing the waveguide system, the initial conditions of the incident wave, and its propagation constant, one can realize a wave field that propagates with a propagation constant belonging to the continuous spectrum while its energy is completely localized within certain bounded regions in space. This phenomenon is known as a \emph{bound state in the continuum(BIC)}.

In this section, we aim to prove the occurrence of a BIC in the waveguide system introduced in this paper with chosen parameters (as in Remark \ref{rmk:3}) and initial data. We will show by theoretical and numerical calculations that the guided wave propagating in the upper and lower waveguides with the same and carefully chosen propagation constant will leak no energy through horizontal waveguides. The propagation constant, however, lies inside the band of horizontal waveguides, which implies the occurrence of BIC.

\subsection{NCME for total system}
Assume that $\phi_+(x,y)=\phi(x,y-D_v)$ and $\phi_-(x,y)=\phi(x,y+D_v)$  
denote the transverse eigen-modes of the upper($+$) and lower($-$) waveguides, representing the field distribution confined in the cross section.  For the total system of $53$ waveguides, we assumed the envelope of the waveguide array is
 $$\psi^t(x,y,z)=c_+(z)\phi_+(x,y)+c_-(z)\phi_-(x,y)+\sum\limits_{m\in\mathbb{Z}^{(25)}}c_m(z)\phi_m(x,y).$$

This is actually the superposition of field in horizontal waveguides we just discussed and the field in two additional waveguides aligned above and below the central waveguide. Now we derive a total NCME for the total $53-$waveguide system like what have done in Section \ref{subsec-3.5}
\begin{equation}\label{TNCME}
i\bm{S}^t\dfrac{\mathrm{d}}{\mathrm{d}z}\bm{C}^t(z)+\bm{K}^t\bm{C}^t(z)=\bm{0},
\end{equation}
where $\bm{C}^t(z)=(c_{-25}(z),\cdots,c_{-1}(z),c_+(z),c_0(z),c_-(z),c_1(z),\cdots,c_{25}(z))^T$, and the matrix $\bm{S}^t$ is given by
\begin{equation}\bm{S}^t=\left(
\begin{array}{ccccccccc}
S^t_{-25,-25}&\cdots &S^t_{-25,-1}&S^t_{-25,+}&S^t_{-25,0}& S^t_{-25,-}&S^t_{-25,1}&\cdots&S^t_{-25,25}\\
\vdots&\cdots &\vdots&\vdots&\vdots& \vdots&\vdots&\cdots&\vdots\\
S^t_{-1,-25}&\cdots &S^t_{-1,-1}&S^t_{-1,+}&S^t_{-1,0}& S^t_{-1,-}&S^t_{-1,1}&\cdots&S^t_{-1,25}\\
S^t_{+,-25}&\cdots &S^t_{+,-1}&S^t_{+,+}&S^t_{+,0}&S^t_{+,-}&S^t_{+,1}&\cdots&S^t_{+,25}\\
S^t_{0,-25}&\cdots &S^t_{0,-1}&S^t_{0,+}&S^t_{0,0}& S^t_{0,-}&S^t_{0,1}&\cdots&S^t_{0,25}\\
S^t_{-,-25}&\cdots &S^t_{-,-1}&S^t_{-,+}&S^t_{-,0}&S^t_{-,-}&S^t_{-,1}&\cdots&S^t_{-,25}\\
\vdots&\cdots &\vdots&\vdots&\vdots& \vdots&\vdots&\cdots&\vdots\\
S^t_{25,-25}&\cdots &S^t_{25,-1}&S^t_{25,+}&S^t_{25,0}& S^t_{25,-}&S^t_{25,1}&\cdots&S^t_{25,25}\\
\end{array}
\right).\end{equation}
$\bm{K}^t$ is defined similarly. The entries of $\bm{S}^t$ and $\bm{K}^t$ are defined as
$$
S^t_{ij}:=\iint_{\Omega}\phi_i(x,y)\phi_j(x,y)\mathrm{d}x\mathrm{d}y,\quad K^t_{ij}:=\iint_{\Omega}\phi_i(x,y)\widehat{H}\phi_j(x,y)\mathrm{d}x\mathrm{d}y
$$
with the operator $\widehat{H}$ defined by \eqref{def_H} and $i,j\in \mathbb{Z}^{(25)}\cup\{+,-\}$. We can use \eqref{formulaSh} to calculate the value of $S_{ij}^t$ with
$$d=\left\{
\begin{aligned}
|i-j| D_h\quad\quad  & \text{for a horizontal waveguide pair}, i,j\in\mathbb{Z}^{(25)}\\
2 D_v\quad\quad  & \text{for a vertical waveguides pair}, \{i,j\} = \{+,-\},\\
 \sqrt{D^2_v+(i D_h)^2}&,\text{ for a horizontal-vertical waveguides pair}, i\in\mathbb{Z}^{(25)}, j\in\{+,-\}.
\end{aligned}\right.$$

\begin{lemma}\label{Indep}Any nonzero function $f\in L^p(\mathbb{R}^n)$ with $p<2n/(n-1)$ has linearly independent translates.(see Edgar~\cite[Corollary~2.7]{edgar1979difference}).
\end{lemma}
\begin{theorem}The matrix $\bm{S}^t$ is invertible.
\end{theorem}
\begin{proof}The matrix $\bm{S}^t \in \mathbb{R}^{53 \times 53}$ has entries given by the overlap integral:
$$S^t_{ij}=\iint_{\Omega}\phi_i(x,y)\phi_j(x,y)\mathrm{d}x\mathrm{d}y,$$ which is called a Gram matrix.
Since $\{\phi_i(x,y)\}_{i\in \mathbb{Z}^{(25)}\cup\{+,-\}}\subseteq L^2(\mathbb{R}^2)$ is linearly independent by Lemma \ref{Indep} when $n=2$ and $\bm{S}^t$ is a Gram matrix. From \cite[p.~441, Theorem~7.2.10]{horn2012matrix} we know that $\bm{S}^t$ is positive definite. Thus, $\bm{S}^t$ is invertible.
\end{proof}
Therefore, the total NCME \eqref{TNCME} could be reformulated as
\begin{equation}\label{TNCME2}
\ii\dfrac{\mathrm{d}}{\mathrm{d}z}\bm{C}^t(z)+\bm{W}^t\bm{C}^t(z)=\bm{0},
\end{equation}
where $\bm{W}^t=(\bm{S}^t)^{-1}\bm{K}^t$.
\begin{definition}\label{def-eigenv}
A solution $\bm{C}^t(z)$ to the equation \eqref{TNCME} is an {\bf eigenmode} if
$$\exists \beta\in\mathbb{R}\text{ and } \bm{C}^t\in\mathbb{C}^{53}\quad \text{ such that }\quad \bm{C}^t(z)=\bm{C}^te^{i\beta z} \text{ and } (\bm{K}^t-\beta\bm{S}^t)\bm{C}^t=\bm{0}.$$
Such $\beta$ are called {\bf eigenvalues} of this system.
\end{definition}

\subsection{Symmetry preserving method}
As the quantity of $S^t_{ij}$ only depends on the distance between two waveguides, for any $i_h\in\mathbb{Z}^{(25)}$, we have
\begin{equation}\label{SymS}S^t_{i_h,+}=S^t_{i_h,-}, \quad S^t_{+,+}=S^t_{-,-} ~~ \text{ and } ~~ S^t_{-,+}=S^t_{+,-}.\end{equation}
\begin{remark}
Here and in what follows, the notation $i_h$ is used to remind that it indexes the horizontal waveguides.
\end{remark}
Now, let's consider $\bm{K}^t$.
\begin{proposition}\label{propKtSym}For any $i_h\in\mathbb{Z}^{(25)}$, it holds that
\begin{equation}\label{SymK}K^t_{i_h,+}=K^t_{i_h,-},\quad K^t_{+,+}=K^t_{-,-}.\end{equation}
\end{proposition}
\begin{proof}
Note that $\phi_{+}(x,-y)=\phi_{-}(x,y)$, $\phi_{i_h}(x,-y)=\phi_{i_h}(x,y)$. By direct calculation we have
$$
\begin{aligned}K^t_{i_h,+}=&\iint_{\Omega}\phi_{i_h}(x,y)\widehat{H}\phi_+(x,y)\mathrm{d}x\mathrm{d}y=\iint_{\Omega}\phi_{i_h}(x,-y)\widehat{H}\phi_+(x,-y)\mathrm{d}x\mathrm{d}y\\
=&\iint_{\Omega}\phi_{i_h}(x,y)\widehat{H}\phi_{-}(x,y)\mathrm{d}x\mathrm{d}y=K^t_{i_h,-}.
\end{aligned}
$$
Another equality follows in the same manner.
\end{proof}
\begin{theorem}\label{thm:bicsolution}
Define
$$
    \beta^t:= \frac{K_{+,+}^t-K_{+,-}^t}{S_{+,+}^t-S_{+,-}^t}.
$$
Consider the system \eqref{TNCME} with initial data
\begin{equation}c_{+}(0) = -c_{-}(0), \qquad c_{i_h}(0)=0 \quad \forall\, i_h\in\mathbb{Z}^{(25)}.\end{equation}
Then a solution $\bm{C}^t(z)$ is given by
\begin{equation}\label{Sol_TNCME}\left\{\begin{aligned}
c_{+}(z) &= c_{+}(0) e^{i\beta^t z}, \\
c_{-}(z) &= c_{-}(0) e^{i\beta^t z},\\
c_{i_h}(z) &= 0, \end{aligned}\right.\end{equation}
where $i_h\in\mathbb{Z}^{(25)}$ and $z\in [0,\infty)$.
\end{theorem}
\begin{proof}
To show \eqref{Sol_TNCME} is solution of \eqref{TNCME}, we need to check
\begin{equation}\label{eq:tmp1}
\ii\left(S^t_{i,+}  \dfrac{\mathrm{d}c_{+}}{\mathrm{d}z}(z) +S^t_{i,-}  \dfrac{\mathrm{d}c_{-}}{\mathrm{d}z}(z) \right)+\left(K^t_{i,+} c_{+}(z) +K^t_{i,-} c_{-}(z)\right)=0, i\in\mathbb{Z}^{(25)}\cup\{+,-\}.
\end{equation}
When $i\in\mathbb{Z}^{(25)}$, the identity follows immediately from \eqref{SymS} and \eqref{SymK}.
When $i\in\{+,-\}$, \eqref{eq:tmp1} is equivalent to
$$
\ii\left(S^t_{\pm,+}-S^t_{\pm,-} \right) \dfrac{\mathrm{d}c_{+}}{\mathrm{d}z}(z)+\left(K^t_{\pm,+}-K^t_{\pm,-} \right)c_{+}(z)=0.
$$
Note that $S^t_{+,+}-S^t_{+,-}=-(S^t_{-,+}-S^t_{-,-})$ and $K^t_{+,+}-K^t_{+,-}=-(K^t_{-,+}-K^t_{-,-})$, thus the equation above is equivalent to
$$
\ii\dfrac{\mathrm{d}c_{+}}{\mathrm{d}z}(z)+\beta^t c_{+}(z)=0.
$$
In conclusion $\bm{C}^t(z)$ defined by \eqref{Sol_TNCME} is a solution to the total NCME \eqref{TNCME}.
\end{proof}
\begin{remark}
Since $\mathbf{W}^t \in \mathbb{C}^{53 \times 53}$ is constant, we can interpret the linear system\eqref{TNCME2}
as an initial value problem on the Banach space $X = \mathbb{C}^{53}$. Defining the linear operator $A := -\ii\mathbf{W}^t$, which is bounded, the system fits the framework of differentiable semigroups. By Pazy~\cite[p.~104, Theorem~1.4]{pazy2012semigroups}, $A$ being the infinitesimal generator of a differentiable semigroup guarantees that the above initial value problem has a unique solution.
\end{remark}
\begin{remark}
The intuition of finding such a solution in Theorem \ref{thm:bicsolution} originates from \cite{Plotnik2011OpticalBIC}. Experimental physicists believe that symmetry plays a crucial role in the emergence of BICs. In the waveguide system studied in this paper, the two additional waveguides attached above and below break the translational symmetry of the system; however, the system still retains symmetry with respect to the $x$-axis. Consequently, a wave that is antisymmetric with respect to the $x$-axis should preserve this symmetry during propagation, that is, the field vanishes identically along the $x$-axis, which exactly matches the characteristics of the BIC mode we aim to find. Using the parameters prescribed in Remark 3, numerical computations show that $\beta^t\approx 795.7056$, which lies precisely within the continuous spectrum (Remark \ref{rmk:9}), thereby confirming that this solution is indeed the desired BIC mode.
\end{remark}

\section{Transition from a perfect BIC to a leaky mode}

From discussion above we could observe that symmetry plays a crucial role in the emergence of BIC. In other words, if the whole system loses symmetry, the BIC is expected to vanish. More precisely, the modes in the vertical waveguides will leak energy to  infinity through horizontal waveguides.

Here we intend to model the paraxial propagation of light in a waveguide array in the same system of linearly coupled modes with broken vertical symmetry. The configuration consists of $N_H = 51$ horizontal waveguides indexed by $m \in \{-25,\ldots,25\}$ and two vertically displaced waveguides, denoted by the symbols $+$ and $-$. In order to break the vertical symmetry, we modify the refractive index of the lower waveguide, i.e. a small positive perturbation $\epsilon^\Theta>0$ is introduced asymmetrically into the refractive indices of the upper and lower waveguides. Specifically, $+\epsilon^\Theta$ is added to the upper waveguide and $-\epsilon^\Theta$ to the lower waveguide. As a result, both the eigenvalue and the coupling coefficient of the vertical waveguide are altered, and hence the coupled-mode equations become significantly more complicated.

In the modified system, the refractive index profile becomes
$$n^\Theta(x,y,z) = n_0 + \Delta n^\Theta(x,y),$$
where
$\Delta n^\Theta(x,y) =\Delta n^\Theta_{+}(x,y)+\Delta n^\Theta_{-}(x,y)+ \sum\limits_{m\in\mathbb{Z}^{(25)}} \Delta n^\Theta_m(x,y),$
and each component  is defined by
$$\Delta n^\Theta_\pm(x,y) =
\begin{cases}
\Delta n\pm\epsilon^\Theta, & (x,y) \in B_a(0,\pm D_v),\\
0, & \text{otherwise},
\end{cases}\quad 
\Delta n^\Theta_m(x,y) =
\begin{cases}
\Delta n, & (x,y) \in B_a(mD_h, 0),\\
0, & \text{otherwise}.
\end{cases}$$
Denote $\phi_i^\Theta(x,y)$ as the mode in each waveguide, $i\in\mathbb{Z}^{(25)}\cup\{+,-\}$. The modes are given by
\begin{equation}\label{phiih}
\phi_{i_h}(x,y)=\left\{\begin{aligned}A J_0(\Lambda \sqrt{(x-i_hD_h)^2+y^2})&,\quad (x,y)\in B_a(i_h D_h, 0),\\
B K_0(\Gamma \sqrt{(x-i_hD_h)^2+y^2})&,\quad (x,y)\in \Omega\setminus B_a(i_hD_h,0),\end{aligned}\right. \quad i_h\in\mathbb{Z}^{(25)},
\end{equation}
\begin{equation}\label{phitheta+}
\phi^\Theta_+(x,y)=\left\{\begin{aligned}A_+^\Theta J_0(\Lambda_+^\Theta \sqrt{x^2+(y-D_v)^2})&,\quad (x,y)\in B_a(0,D_v),\\B_+^\Theta K_0(\Gamma_+^\Theta \sqrt{x^2+(y-D_v)^2})&,\quad (x,y)\in \Omega\setminus B_a(0,D_v),\end{aligned}\right.
\end{equation}
\begin{equation}\label{phitheta-}
\phi^\Theta_-(x,y)=\left\{\begin{aligned}A_-^\Theta J_0(\Lambda_-^\Theta \sqrt{x^2+(y+D_v)^2})&,\quad (x,y)\in B_a(0,-D_v),\\B_-^\Theta K_0(\Gamma_-^\Theta \sqrt{x^2+(y+D_v)^2})&,\quad (x,y)\in \Omega\setminus B_a(0,-D_v),\end{aligned}\right.
\end{equation}
where \begin{equation}\label{lambdagammatheta+}
\Lambda^\Theta_+
=\sqrt{\dfrac{8\pi^2n_0(\Delta n+\epsilon^\Theta)}{\lambda^2}-\dfrac{4\pi n_0\beta^{\Theta,+}_0}{\lambda}},
\quad
\Gamma^\Theta_+
=\sqrt{\dfrac{4\pi n_0\beta^{\Theta,+}_0}{\lambda}},
\end{equation}
\begin{equation}\label{lambdagammatheta-}
\Lambda^\Theta_-
=\sqrt{\dfrac{8\pi^2n_0(\Delta n-\epsilon^\Theta)}{\lambda^2}-\dfrac{4\pi n_0\beta^{\Theta,-}_0}{\lambda}},
\quad
\Gamma^\Theta_-
=\sqrt{\dfrac{4\pi n_0\beta^{\Theta,-}_0}{\lambda}},
\end{equation}
\begin{equation}\label{formulaABtheta}A^\Theta_+ = \left[ \pi a^2 \left( J^2_1(\Lambda^\Theta_+ a) + J^2_0(\Lambda^\Theta_+  a) \right) + \pi a^2 \left(\dfrac{J_0(\Lambda^\Theta_+  a)}{K_0(\Gamma^\Theta_+  a)}\right)^2(K^2_1(\Gamma^\Theta_+  a)-K^2_0(\Gamma^\Theta_+  a))\right]^{-1/2},
\quad
B^\Theta_+  = A^\Theta_+  \frac{J_0(\Lambda^\Theta_+  a)}{K_0(\Gamma^\Theta_+  a)}.
\end{equation}
\begin{equation}A^\Theta_{-} = \left[ \pi a^2 \left( J^2_1(\Lambda^\Theta_{-}  a) + J^2_0(\Lambda^\Theta_{-}  a) \right) +
\pi a^2 \left(\dfrac{J_0(\Lambda^\Theta_{-}  a)}{K_0(\Gamma^\Theta_{-}  a)}\right)^2(K^2_1(\Gamma^\Theta_{-}  a)-K^2_0(\Gamma^\Theta_{-}  a))\right]^{-1/2},
\quad
B^\Theta_{-}  = A^\Theta_{-} \frac{J_0(\Lambda^\Theta_{-}  a)}{K_0(\Gamma^\Theta_{-}  a)}.\end{equation}
Similar to \eqref{governbeta}, $\beta^{\Theta,+}_0$ and $\beta^{\Theta,-}_0$ are governed by
\begin{equation}\label{governbeta+-}\dfrac{\Lambda^\Theta_+ J_1(\Lambda^\Theta_+ a)}{J_0(\Lambda^\Theta_+ a)} = \dfrac{\Gamma^\Theta_+ K_1(\Gamma^\Theta_+ a)}{K_0(\Gamma^\Theta_+ a)},\quad \dfrac{\Lambda^\Theta_- J_1(\Lambda^\Theta_- a)}{J_0(\Lambda^\Theta_- a)} = \dfrac{\Gamma^\Theta_- K_1(\Gamma^\Theta_- a)}{K_0(\Gamma^\Theta_- a)}.\end{equation}

We consider the evolution of the envelope along the propagation direction $z$ as the linear superposition of modes in each waveguide, i.e.
 \begin{equation}\label{psitheta}\psi^t_\Theta(x,y,z)=c_+(z)\phi^\Theta_+(x,y)+c_-(z)\phi^\Theta_{-}(x,y)+\sum\limits_{i_h\in\mathbb{Z}^{(25)}}c_{i_h}(z)\phi_{i_h}(x,y).\end{equation}
$\psi^t_\Theta$ is governed by the following modified NCME for the $53$-waveguide system:
\begin{equation}\label{MTNCME}
\ii\bm{S}^t_\Theta\dfrac{\mathrm{d}}{\mathrm{d}z}\bm{C}^t(z)+\bm{K}^t_\Theta\bm{C}^t(z)=\bm{0},
\end{equation}
where $\bm{C}^t(z)=(c_{-25}(z),\cdots,c_{-1}(z),c_+(z),c_0(z),c_-(z),c_1(z),\cdots,c_{25}(z))^{\mathsf{T}}\in \mathbb{C}^{53}$, and the entries of matrices $\bm{S}^t_\Theta$ and $ \bm{K}^t_\Theta$ are
\begin{align*}
S^\Theta_{ij}&=\iint_{\Omega}\phi_i(x,y)\phi_j(x,y)\mathrm{d}x\mathrm{d}y, \quad \kappa^\Theta_{ij}= \iint_{\Omega}\phi_i(x,y)\sum\limits_{l\neq j}\dfrac{k\Delta n^\Theta_l(x,y)}{n_0}\phi_j(x,y)\mathrm{d}x \mathrm{d}y, \\
K^\Theta_{ij}&=\iint_{\Omega}\phi_i(x,y)\widehat{H}^\Theta\phi_j(x,y)\mathrm{d}x\mathrm{d}y,
\end{align*}
for $i,j\in \mathbb{Z}^{(25)}\cup\{+,-\}$. The operator $\widehat{H}^\Theta$ defined by
\begin{equation}\label{def_HTheta}
\widehat{H}^\Theta=\left(\dfrac{1}{2k}(\partial_x^2+\partial_y^2)+\dfrac{k\Delta n^\Theta(x,y)}{n_0}\right).
\end{equation}

\subsection{The formulas for the matrices \texorpdfstring{$\bm{S}^t$}{St} and  \texorpdfstring{$\bm{\kappa}^t$}{kappat} in the broken symmetry case}

\begin{proposition}\label{SKThetasym}The matrices $\bm{S}^t_\Theta$ and  $\bm{K}^t_\Theta$ are symmetric. However, the symmetry of $\bm{\kappa}^t_\Theta$ is broken. Fortunately, the entries of $\bm{\kappa}^t_\Theta$ corresponding to the two horizontal waveguides still satisfy the symmetry.
\end{proposition}
\begin{proof}From the definition alone, one can already tell that the matrix $\bm{S}^t_\Theta$ is symmetric.
Applying Lemma~\ref{Hselfadjoint}, the operator $\widehat{H}^\Theta$ is self-adjoint, and hence $\bm{K}^t_\Theta$ is also symmetric. Note that the transverse mode functions are eigenfunctions of the operator $$\widehat{H}^\Theta_i=\left(\dfrac{1}{2k}(\partial_x^2+\partial_y^2)+\dfrac{k\Delta n^\Theta_i (x,y)}{n_0}\right),$$
for the vertical waveguides, there holds that
$$K^\Theta_{i_h,+}=\beta_0^{\Theta,+}S^\Theta_{i_h,+}+\kappa^\Theta_{i_h,+},\quad
K^\Theta_{+,i_h}=\beta_0 S^\Theta_{+,i_h}+\kappa^\Theta_{+,i_h},$$
$$K^\Theta_{i_h,-}=\beta_0^{\Theta,-}S^\Theta_{i_h,-}+\kappa^\Theta_{i_h,-},\quad
K^\Theta_{-,i_h}=\beta_0 S^\Theta_{-,i_h}+\kappa^\Theta_{-,i_h}.$$
This breaks the symmetry of $\bm{\kappa}^t_\Theta$ that the others maintain.  However, for two horizontal waveguides we still have $\kappa^\Theta_{i_h,j_h}=\kappa^\Theta_{j_h,i_h}$.
\end{proof}
\begin{theorem}\label{thm-Sthetaformula} For  pairs of horizontal waveguides, the entries of $\bm{S}^t_\Theta$ are given by \eqref{formulaSh}. For horizontal and vertical waveguides pairs, they are given by
\begin{equation}S^\Theta_{+,+}=S^\Theta_{-,-}=1,
\end{equation}
\begin{equation}\label{Stheta+}\begin{aligned}
S^\Theta_{i_h,+}&= 2\pi A^\Theta_{+} B K_0(\Gamma d_{i_h}) \cdot \frac{a}{(\Lambda^\Theta_{+})^2 + (\Gamma)^2} \left[ \Lambda^\Theta_{+} I_0(\Gamma a) J_1(\Lambda^\Theta_{+} a) + \Gamma J_0(\Lambda^\Theta_{+} a) I_1(\Gamma a) \right]\\
& + 2\pi A B^\Theta_{+} K_0(\Gamma^\Theta_{+} d_{i_h}) \cdot \frac{a}{\Lambda^2 + (\Gamma^\Theta_{+})^2} \left[ \Lambda I_0(\Gamma^\Theta_{+} a) J_1(\Lambda a) + \Gamma^\Theta_{+} J_0(\Lambda a) I_1(\Gamma^\Theta_{+} a) \right]\\
&+ \dfrac{2\pi B B^\Theta_+ a }{ (\Gamma^\Theta_+)^2-\Gamma^2 } \left[ K_0(\Gamma d_{i_h}) \left( \Gamma^\Theta_+ I_0(\Gamma a) K_1(\Gamma^\Theta_+ a) + \Gamma I_1(\Gamma a) K_0(\Gamma^\Theta_+ a) \right)\right.  \\
& \left.-K_0(\Gamma^\Theta_+ d_{i_h}) \left( \Gamma I_0(\Gamma^\Theta_+ a) K_1(\Gamma a) + \Gamma^\Theta_+ I_1(\Gamma^\Theta_+ a) K_0(\Gamma a) \right)\right],
\end{aligned}
\end{equation}
\begin{equation}\label{Stheta-}\begin{aligned}
S^\Theta_{i_h,-}&= 2\pi A^\Theta_{-} B K_0(\Gamma d_{i_h}) \cdot \frac{a}{(\Lambda^\Theta_{-})^2 + (\Gamma)^2} \left[ \Lambda^\Theta_{-} I_0(\Gamma a) J_1(\Lambda^\Theta_{-} a) + \Gamma J_0(\Lambda^\Theta_{-} a) I_1(\Gamma a) \right]\\
& + 2\pi A B^\Theta_{-} K_0(\Gamma^\Theta_{-} d_{i_h}) \cdot \frac{a}{\Lambda^2 + (\Gamma^\Theta_{-})^2} \left[ \Lambda I_0(\Gamma^\Theta_{-} a) J_1(\Lambda a) + \Gamma^\Theta_{-} J_0(\Lambda a) I_1(\Gamma^\Theta_{-} a) \right]\\
&+ \dfrac{2\pi B B^\Theta_{-} a }{\Gamma^2 - (\Gamma^\Theta_{-})^2} \left[ K_0(\Gamma^\Theta_{-} d_{i_h}) \left( \Gamma I_0(\Gamma^\Theta_{-} a) K_1(\Gamma a) + \Gamma^\Theta_{-} I_1(\Gamma^\Theta_{-} a) K_0(\Gamma a) \right)\right.\\
& \left.- K_0(\Gamma d_{i_h}) \left( \Gamma^\Theta_{-} I_0(\Gamma a) K_1(\Gamma^\Theta_{-} a) + \Gamma I_1(\Gamma a) K_0(\Gamma^\Theta_{-} a) \right) \right],
\end{aligned}
\end{equation}
\begin{equation}\label{Stheta+-}\begin{aligned}
S^\Theta_{+,-}
&= 2\pi A^\Theta_{-} B^\Theta_+ K_0(\Gamma^\Theta_+ 2D_v) \cdot \dfrac{a}{(\Lambda^\Theta_{-})^2 + (\Gamma^\Theta_+)^2} \left[ \Lambda^\Theta_{-} I_0(\Gamma^\Theta_+ a) J_1(\Lambda^\Theta_{-} a) + \Gamma^\Theta_+ J_0(\Lambda^\Theta_{-} a) I_1(\Gamma^\Theta_+ a) \right]\\
& + 2\pi A^\Theta_+ B^\Theta_{-} K_0(\Gamma^\Theta_{-} 2 D_v) \cdot \dfrac{a}{(\Lambda^\Theta_+)^2 + (\Gamma^\Theta_{-})^2} \left[ \Lambda^\Theta_+ I_0(\Gamma^\Theta_{-} a) J_1(\Lambda^\Theta_+ a) + \Gamma^\Theta_{-} J_0(\Lambda^\Theta_+ a) I_1(\Gamma^\Theta_{-} a) \right]\\
&+ \dfrac{2\pi B^\Theta_+ B^\Theta_{-} a }{(\Gamma^\Theta_+)^2 - (\Gamma^\Theta_{-})^2} \left[ K_0(\Gamma^\Theta_{-} 2D_v) \left( \Gamma^\Theta_+ I_0(\Gamma^\Theta_{-} a) K_1(\Gamma^\Theta_+ a) + \Gamma^\Theta_{-} I_1(\Gamma^\Theta_{-} a) K_0(\Gamma^\Theta_+ a) \right)\right.\\
& \left.- K_0(\Gamma^\Theta_+ 2D_v) \left( \Gamma^\Theta_{-} I_0(\Gamma^\Theta_+ a) K_1(\Gamma^\Theta_{-} a) + \Gamma^\Theta_+ I_1(\Gamma^\Theta_+ a) K_0(\Gamma^\Theta_{-} a) \right) \right],
\end{aligned}
\end{equation}
where $d_{i_h}=\sqrt{(i_hD_h)^2+D_v^2}$ is the distance between two waveguides.
\end{theorem}

\begin{proof}
Since the functions $\phi^\Theta_+(x,y)$ and $\phi_{i_h}(x,y)$ are piecewise constant, we separate the plane $\Omega$ into three subsets:
$\Omega_1=B_a(0,D_v),\Omega_2=B_a(i_hD_h,0)$ and $\Omega_3=\Omega\setminus(\Omega_1\cup\Omega_2)$. Using polar coordinates for each subset, one can obtain
$$\begin{aligned}
S^\Theta_{i_h,+}
&=\iint_{\Omega_1}\phi^\Theta_{+}(x,y)\phi_{i_h}(x,y)\mathrm{d}x\mathrm{d}y+\iint_{\Omega_2}\phi^\Theta_{+}(x,y)\phi_{i_h}(x,y)\mathrm{d}x\mathrm{d}y+\iint_{\Omega_3}\phi^\Theta_{+}(x,y)\phi_{i_h}(x,y)\mathrm{d}x\mathrm{d}y\\
&=:I^\Theta_1+I^\Theta_2+I^\Theta_3,
\end{aligned}
$$
where
\begin{align*}
I^\Theta_1 &= A^\Theta_+ B \int_0^{2\pi} \mathrm{d}\theta\int_0^a J_0(\Lambda^\Theta_+ r) K_0(\Gamma \sqrt{r^2 + d_{i_h}^2 + 2 d_{i_h} r \cos \theta}) r \mathrm{d}r, \\
I^\Theta_2 &= A B^\Theta_+ \int_0^{2\pi} \mathrm{d}\theta\int_0^a J_0(\Lambda r) K_0(\Gamma^\Theta_+ \sqrt{r^2 + d_{i_h}^2 - 2 d_{i_h} r \cos \theta}) r \mathrm{d}r, \\
I^\Theta_3&=B\cdot B^\Theta_+\iint_{\Omega_3}  K_0(\Gamma \sqrt{(x - i_h D_h)^2 + y^2})K_0(\Gamma^\Theta_+ \sqrt{x^2 + (y-D_v)^2})  \mathrm{d}x \mathrm{d}y.
\end{align*}
To evaluate the integral $I_1$ and $I_2$, one applies the addition theorem \ref{Additiontheorem}  and gets
$$I^\Theta_1 = A^\Theta_+ B \cdot 2\pi K_0(\Gamma d_{i_h}) \int_0^a J_0(\Lambda^\Theta_+ r) I_0(\Gamma r) r\mathrm{d}r,$$
$$I^\Theta_2 = AB^\Theta_+  \cdot 2\pi K_0(\Gamma^\Theta_+  d_{i_h}) \int_0^a J_0(\Lambda r) I_0(\Gamma^\Theta_+  r) r\mathrm{d}r.$$
The integral above matches the formula \eqref{eq:bessel4}, then
$$I^\Theta_1 = 2\pi A^\Theta_{+} B K_0(\Gamma d_{i_h}) \cdot \frac{a}{(\Lambda^\Theta_{+})^2 + \Gamma^2} \left[ \Lambda^\Theta_{+} I_0(\Gamma a) J_1(\Lambda^\Theta_{+} a) + \Gamma J_0(\Lambda^\Theta_{+} a) I_1(\Gamma a) \right].$$
$$I^\Theta_2 = 2\pi A B^\Theta_{+} K_0(\Gamma^\Theta_{+} d_{i_h}) \cdot \frac{a}{\Lambda^2 + (\Gamma^\Theta_{+})^2} \left[ \Lambda I_0(\Gamma^\Theta_{+} a) J_1(\Lambda a) + \Gamma^\Theta_{+} J_0(\Lambda a) I_1(\Gamma^\Theta_{+} a) \right].$$
Regarding the integral $I^\Theta_3$, the domain $\Omega_3$ is quite complex. So one can use cut-and-paste method as following:
$$I^\Theta_3=I_3^{\Theta,total}-I_3^{\Theta, disk,1}-I_3^{\Theta, disk,2}$$
where
\begin{align*}
I_3^{\Theta, total}&:= B\cdot B^\Theta_+\iint_{\Omega}  K_0(\Gamma \sqrt{(x - i_h D_h)^2 + y^2})K_0(\Gamma^\Theta_+ \sqrt{x^2 + (y-D_v)^2})  \mathrm{d}x \mathrm{d}y, \\
I_3^{\Theta, disk,1}&:= B\cdot B^\Theta_+\iint_{B_a(0,D_v)}  K_0(\Gamma \sqrt{(x - i_h D_h)^2 + y^2})K_0(\Gamma^\Theta_+ \sqrt{x^2 + (y-D_v)^2})  \mathrm{d}x \mathrm{d}y,\\
I_3^{\Theta, disk,2}&:= B\cdot B^\Theta_+\iint_{B_a(i_hD_h,0)}  K_0(\Gamma \sqrt{(x - i_h D_h)^2 + y^2})K_0(\Gamma^\Theta_+ \sqrt{x^2 + (y-D_v)^2})  \mathrm{d}x \mathrm{d}y.
\end{align*}
In polar coordinates,
$$I_3^{\Theta, total} = B\cdot B^\Theta_+ \int_0^{2\pi} \mathrm{d}\theta\int_0^\infty K_0(\Gamma r) K_0(\Gamma^\Theta_+ \sqrt{r^2 + d_{i_h}^2 - 2 r d_{i_h} \cos \theta}) r\mathrm{d}r .$$
Applying the addition theorem for $K_0$ again yields
$$\begin{aligned}
I_3^{\Theta, total} &= B\cdot B^\Theta_+ \int_0^{2\pi}\mathrm{d}\theta  \int_0^{d_{i_h}} K_0(\Gamma r) \sum\limits_{m=-\infty}^{+\infty} I_m(\Gamma^\Theta_+ r) K_m(\Gamma^\Theta_+ d_{i_h}) \cos(m \theta) r \mathrm{d}r \\
&+B\cdot B^\Theta_+ \int_0^{2\pi}\mathrm{d}\theta \int_{d_{i_h}}^\infty K_0(\Gamma r) \sum\limits_{m=-\infty}^{+\infty} I_m(\Gamma^\Theta_+ d_{i_h}) K_m(\Gamma^\Theta_+ r) \cos(m \theta) r \mathrm{d}r\\
&= 2\pi B\cdot B^\Theta_+ \left[ K_0(\Gamma^\Theta_+ d_{i_h}) \int_0^{d_{i_h}} K_0(\Gamma r) I_0(\Gamma^\Theta_+ r) r \mathrm{d}r + I_0(\Gamma^\Theta_+ d_{i_h}) \int_{d_{i_h}}^\infty K_0(\Gamma r)K_0(\Gamma^\Theta_+ r) r \mathrm{d}r\right].\end{aligned}$$
Then the formulas \eqref{int0x2K0I0} and \eqref{intx1inftyK0K0} give
$$\begin{aligned}I_3^{\Theta, total}
=\dfrac{2\pi B\cdot B^\Theta_+}{\Gamma^2-(\Gamma^\Theta_+)^2}
&\left[ K_0(\Gamma^\Theta_+ d_{i_h})-d_{i_h}K_0(\Gamma^\Theta_+ d_{i_h})
\left(\Gamma^\Theta_+ I_1(\Gamma^\Theta_+ d_{i_h})K_0(\Gamma d_{i_h})+\Gamma I_0(\Gamma^\Theta_+ d_{i_h})K_1(\Gamma d_{i_h})\right)\right.\\
&\left.+d_{i_h}I_0(\Gamma^\Theta_+ d_{i_h}) \left(\Gamma K_1(\Gamma d_{i_h})K_0(\Gamma^\Theta_+ d_{i_h})-\Gamma^\Theta_+ K_0(\Gamma d_{i_h})K_1(\Gamma^\Theta_+d_{i_h})\right)\right]\\
=\dfrac{2\pi B\cdot B^\Theta_+}{\Gamma^2-(\Gamma^\Theta_+)^2}&\left[ K_0(\Gamma^\Theta_+ d_{i_h}) - \Gamma^\Theta_+ d_{i_h} K_0(\Gamma d_{i_h}) \left( I_0(\Gamma^\Theta_+ d_{i_h}) K_1(\Gamma^\Theta_+ d_{i_h}) + I_1(\Gamma^\Theta_+ d_{i_h}) K_0(\Gamma^\Theta_+ d_{i_h}) \right)\right].
\end{aligned}
$$
Using the known Wronskian identity for modified Bessel functions $I_0(x) K_1(x) + I_1(x) K_0(x) = \dfrac{1}{x}$, we have
$$I_3^{\Theta, total}
=\dfrac{2\pi B\cdot B^\Theta_+}{\Gamma^2-(\Gamma^\Theta_+)^2}\left[ K_0(\Gamma^\Theta_+ d_{i_h}) - \Gamma^\Theta_+ d_{i_h} K_0(\Gamma d_{i_h}) \cdot \frac{1}{\Gamma^\Theta_+ d_{i_h}}\right] = \dfrac{2\pi B\cdot B^\Theta_+}{\Gamma^2-(\Gamma^\Theta_+)^2}[K_0(\Gamma^\Theta_+ d_{i_h}) - K_0(\Gamma d_{i_h})].$$

Using the same method of calculation for $I^\Theta_1$ and $I^\Theta_2$ , one can obtain
$$I_3^{\Theta, disk,1}= 2\pi B\cdot B^\Theta_+K_0(\Gamma d_{i_h})\int_0^a K_0(\Gamma^\Theta_+ r)I_0(\Gamma r)r\mathrm{d}r,$$
$$I_3^{\Theta, disk,2}= 2\pi B\cdot B^\Theta_+K_0(\Gamma^\Theta_+ d_{i_h})\int_0^a K_0(\Gamma r)I_0(\Gamma^\Theta_+ r)r\mathrm{d}r.$$
Applying the formulas \eqref{int0x2K0I0}, it gives that
\begin{align*}
I_3^{\Theta, disk,1}&=
\dfrac{2\pi B\cdot B^\Theta_+}{(\Gamma^\Theta_+)^2-\Gamma^2} \left[K_0(\Gamma d_{i_h})-aK_0(\Gamma d_{i_h})\left(\Gamma^\Theta_+ I_0(\Gamma a)K_1(\Gamma^\Theta_+ a)+\Gamma I_1(\Gamma a)K_0(\Gamma^\Theta_+ a)\right)\right], \\
I_3^{\Theta, disk,2}&=
\dfrac{2\pi B\cdot B^\Theta_+}{\Gamma^2-(\Gamma^\Theta_+)^2} \left[K_0(\Gamma^\Theta_+ d_{i_h})-aK_0(\Gamma^\Theta_+ d_{i_h})\left(\Gamma I_0(\Gamma^\Theta_+ a)K_1(\Gamma a)+\Gamma I_1(\Gamma^\Theta_+ a)K_0(\Gamma a)\right)\right].
\end{align*}
Hence,
$$\begin{aligned}I^\Theta_1 + I^\Theta_2 + I^\Theta_3 &= 2\pi A^\Theta_{+} B K_0(\Gamma d_{i_h}) \cdot \frac{a}{(\Lambda^\Theta_{+})^2 + \Gamma^2} \left[ \Lambda^\Theta_{+} I_0(\Gamma a) J_1(\Lambda^\Theta_{+} a) + \Gamma J_0(\Lambda^\Theta_{+} a) I_1(\Gamma a) \right]\\
& + 2\pi A B^\Theta_{+} K_0(\Gamma^\Theta_{+} d_{i_h}) \cdot \frac{a}{\Lambda^2 + (\Gamma^\Theta_{+})^2} \left[ \Lambda I_0(\Gamma^\Theta_{+} a) J_1(\Lambda a) + \Gamma^\Theta_{+} J_0(\Lambda a) I_1(\Gamma^\Theta_{+} a) \right]\\
&+ \dfrac{2\pi B B^\Theta_+ a }{ (\Gamma^\Theta_+)^2-\Gamma^2} \left[K_0(\Gamma d_{i_h}) \left( \Gamma^\Theta_+ I_0(\Gamma a) K_1(\Gamma^\Theta_+ a) + \Gamma I_1(\Gamma a) K_0(\Gamma^\Theta_+ a) \right)\right. \\
& \left.- K_0(\Gamma^\Theta_+ d_{i_h}) \left( \Gamma I_0(\Gamma^\Theta_+ a) K_1(\Gamma a) + \Gamma^\Theta_+ I_1(\Gamma^\Theta_+ a) K_0(\Gamma a) \right) \right].
\end{aligned}$$
$S^\Theta_{i_h,-}$ and $S^\Theta_{+,-}$ can be calculated in the same way, which are the overlap integrals related to corresponding waveguide pairs. For brevity, we omit the calculation details.
\end{proof}

Now, we start to establish the formula of the entries of $\bm{K}^t_\Theta$. According to the previous results about $\bm{\kappa}^t_\Theta$ in this section, we only need to modify the coefficients of the integrals whose domain is $B_a(0,\pm D_v)$ and the parameters $\Gamma$ and $\Lambda$ in the transverse mode functions for the vertical waveguides. By Proposition~\ref{SKThetasym}, the matrix $\bm{K}^t_\Theta$ is symmetric. Consequently,  we need to compute the following entries of $\bm{K}^t_\Theta$:
$$K^\Theta_{i_h,+}, K^\Theta_{i_h,-}, K^\Theta_{i_h,j_h}, K^\Theta_{+,+}, K^\Theta_{-,-}, K^\Theta_{-,+}$$
where $i_h, j_h\in\mathbb{Z}^{(25)}\text{ and } i_h\leqslant j_h.$
Since  we have known the matrix $\bm{S}^t_\Theta$, it suffices to calculate the following entries of $\bm{\kappa}^t_\Theta$:
$$\kappa^\Theta_{i_h,+}, \kappa^\Theta_{i_h,-}, \kappa^\Theta_{i_h,j_h}, \kappa^\Theta_{+,+}, \kappa^\Theta_{-,-}, \kappa^\Theta_{-,+}.$$
where $i_h, j_h\in\mathbb{Z}^{(25)}\text{ and } i_h\leqslant j_h.$

We only state the results and omit the details of the proof, since the argument
follows verbatim that of the previous two subsections, with only a change of parameters.
For clarity, the method of computation with the new parameters is carried out explicitly in
Theorem~\ref{thm-Sthetaformula}.

Define $R_m := \sqrt{(m D_h)^2 + D_v^2}$, and $\alpha_m\in[0,2\pi)$, $m\in\mathbb{Z}^{(25)}$ satisfy $\cos \alpha_m = \dfrac{ - m D_h }{ R_m }$ and $\sin \alpha_m = \dfrac{ D_v }{ R_m }$.

\begin{proposition}\label{prop:63}
$\kappa^\Theta_{i_h,j_h}=\kappa^t_{ij}$, namely,
\begin{equation}\label{kappathetaij}\kappa^\Theta_{i_h,j_h}=\dfrac{k\Delta n}{n_0} \left(\sum\limits_{m\in\mathbb{Z}^{(25)}\setminus \{j\}}I_{m,i,j}~~ +2 I^v_{i,j}\right).\end{equation}

where
\begin{equation}\label{Imij}\footnotesize
I_{m,i,j}=
\left\{\begin{aligned}
\pi B^2 a^2 \bigg[ K_0(\Gamma \vert m-i\vert D_h) K_0(\Gamma \vert m-j\vert D_h)(I_0^2(\Gamma a)-I_1^2(\Gamma a))
+ 2\sum_{q=1}^\infty K_q(\Gamma \vert m-i\vert  D_h) K_q(\Gamma \vert m-j\vert D_h)(I_q^2(\Gamma a)-I_{q-1}(\Gamma a)I_{q+1}(\Gamma a) )\bigg]
&,(m-i)(m-j)>0\\
\pi B^2 a^2\bigg[K_0(\Gamma (m-i) D_h)K_0(\Gamma(j-m)D_h)(I_0^2(\Gamma a)-I_1^2(\Gamma a)) +2\sum\limits_{q=1}^\infty(-1)^qK_q(\Gamma (m-i) D_h)K_q(\Gamma(j-m)D_h)(I_q^2(\Gamma a)-I_{q-1}(\Gamma a)I_{q+1}(\Gamma a))\bigg]&,i<m<j\\
\dfrac{2\pi ABa}{\Gamma^2+\Lambda^2}K_0(\Gamma \vert j-i\vert D_h)(\Lambda I_0(\Gamma a)J_1(\Lambda a)+\Gamma J_0(\Lambda a)I_1(\Gamma a))
\quad\quad\quad\quad\quad\quad\quad\quad\quad\quad\quad\quad\quad\quad
&,m=i.
\end{aligned}\right.
\end{equation}

and
\begin{equation}\label{Ivij}\begin{aligned} I^v_{i,j}=&\pi B^2a^2K_0(\Gamma R_i)K_0(\Gamma R_j)(I_0^2(\Gamma a)-I_1^2(\Gamma a))\\
&+2\pi B^2 a^2\sum\limits_{q=1}^\infty K_q(\Gamma R_i)K_q(\Gamma R_j)\cos(q(\alpha_i-\alpha_j))(I_q^2(\Gamma a)-I_{q-1}(\Gamma a)I_{q+1}(\Gamma a)).\end{aligned}\end{equation}
\end{proposition}

\begin{proposition}\label{prop:64}$\kappa^\Theta_{+,+}$ can be expressed as
\begin{equation}\label{kappatheta++}\kappa^\Theta_{+,+}=\dfrac{k\Delta n}{n_0}\sum\limits_{m\in\mathbb{Z}^{(25)}}I^{m,\Theta}_{+,+}+\dfrac{k(\Delta n-\epsilon^\Theta)}{n_0}I^{v,\Theta}_{+,+},
\end{equation}

where
 \begin{equation}\label{Ithetam++} \begin{aligned} I^{m,\Theta}_{+,+}=&\pi (B^\Theta_+)^2 a^2 K^2_0(\Gamma^\Theta_+ R_m)(I_0^2(\Gamma^\Theta_+ a)-I_1^2(\Gamma^\Theta_+ a))\\
&+2\pi (B^\Theta_+)^2 a^2\sum\limits_{q=1}^\infty K^2_q(\Gamma^\Theta_+ R_m)(I_q^2(\Gamma^\Theta_+ a)-I_{q-1}(\Gamma^\Theta_+ a)I_{q+1}(\Gamma^\Theta_+ a)).
\end{aligned}
\end{equation}

and
\begin{equation}\label{Ithetav++}\begin{aligned}
I^{v,\Theta}_{+,+} =& \pi (B^\Theta_+)^2a^2  K^2_0(2\Gamma^\Theta_+ D_v) (I^2_0(\Gamma^\Theta_+ a)-I^2_1(\Gamma^\Theta_+ a))\\
&+ 2\pi (B^\Theta_+)^2a^2 \sum_{q=1}^\infty K^2_q(2\Gamma^\Theta_+ D_v) (I_q^2(\Gamma^\Theta_+ a)-I_{q-1}(\Gamma^\Theta_+ a)I_{q+1}(\Gamma^\Theta_+ a)).
\end{aligned}\end{equation}
\end{proposition}

\begin{proposition}\label{prop:65}
$\kappa^\Theta_{-,-}$ can be expressed as
\begin{equation}\label{kappatheta--}\kappa^\Theta_{-,-}=\dfrac{k\Delta n}{n_0}\sum\limits_{m\in\mathbb{Z}^{(25)}}I^{m,\Theta}_{-,-}+\dfrac{k(\Delta n+\epsilon^\Theta)}{n_0}I^{v,\Theta}_{-,-},
\end{equation}

where
\begin{equation}\label{Ithetam--}\begin{aligned} I^{m,\Theta}_{-,-}=&\pi (B^\Theta_-)^2 a^2 K^2_0(\Gamma^\Theta_{-} R_m)(I_0^2(\Gamma^\Theta_- a)-I_1^2(\Gamma^\Theta_{-} a))\\
&+2\pi (B^\Theta_{-})^2 a^2\sum\limits_{q=1}^\infty K^2_q(\Gamma^\Theta_- R_m)(I_q^2(\Gamma^\Theta_- a)-I_{q-1}(\Gamma^\Theta_{-} a)I_{q+1}(\Gamma^\Theta_{-} a)).\end{aligned}\end{equation}

and
\begin{equation}\label{Ithetav--}
\begin{aligned}
I^{v,\Theta}_{-,-} =& \pi (B^\Theta_-)^2a^2  K^2_0(2\Gamma^\Theta_- D_v) (I^2_0(\Gamma^\Theta_- a)-I^2_1(\Gamma^\Theta_- a))\\
&+ 2\pi (B^\Theta_-)^2a^2 \sum_{q=1}^\infty K^2_q(2\Gamma^\Theta_- D_v) (I_q^2(\Gamma^\Theta_- a)-I_{q-1}(\Gamma^\Theta_- a)I_{q+1}(\Gamma^\Theta_- a)).
\end{aligned}
\end{equation}
\end{proposition}

\begin{proposition}\label{prop:66}
$\kappa^\Theta_{-,+}$ can be expressed as
\begin{equation}\label{kappatheta-+}\kappa^\Theta_{-,+}=\dfrac{k\Delta n}{n_0}\sum\limits_{m\in\mathbb{Z}^{(25)}}I^{m,\Theta}_{-,+}+\dfrac{k(\Delta n-\epsilon^\Theta)}{n_0}I^{v,\Theta}_{-,+},\end{equation}

where
\begin{equation}\label{Ithetam-+}
\begin{aligned}
 I^{m,\Theta}_{-,+}=&\dfrac{2\pi B^\Theta_+B^\Theta_{-}a}{(\Gamma^\Theta_+)^2-(\Gamma^\Theta_{-})^2}K_0(\Gamma^\Theta_+ R_m)K_0(\Gamma^\Theta_{-} R_m)
(\Gamma^\Theta_+I_0(\Gamma^\Theta_{-}a)I_1(\Gamma^\Theta_+ a)-\Gamma^\Theta_{-}I_1(\Gamma^\Theta_{-} a)I_0(\Gamma^\Theta_+ a))\\
&+\dfrac{4\pi B^\Theta_+B^\Theta_{-} a}{(\Gamma^\Theta_+)^2-(\Gamma^\Theta_{-})^2 }\sum\limits_{q=1}^\infty K_q(\Gamma^\Theta_+ R_m)K_q(\Gamma^\Theta_{-} R_m)\cos(2q\alpha_m)(\Gamma^\Theta_+I_q(\Gamma^\Theta_{-}a)I_{q-1}(\Gamma^\Theta_+ a)-\Gamma^\Theta_{-}I_{q-1}(\Gamma^\Theta_{-} a)I_q(\Gamma^\Theta_+ a)).
\end{aligned}\end{equation}

and
\begin{equation}\label{Ithetav-+}
I^{v,\Theta}_{-,+} = \dfrac{2\pi A^\Theta_{-}B^\Theta_+ a}{(\Lambda^\Theta_{-})^2+(\Gamma^\Theta_+)^2}K_0(2\Gamma^\Theta_+ D_v)
(\Lambda^\Theta_{-}I_0(\Gamma^\Theta_+ a)J_1(\Lambda^\Theta_{-}a)+\Gamma^\Theta_+ J_0(\Lambda^\Theta_{-}a)I_1(\Gamma^\Theta_+ a)).
\end{equation}
\end{proposition}

\begin{proposition}\label{prop:67}
$\kappa^\Theta_{i_h,+}$ can be expressed as
\begin{equation}\label{kappathetai+}
\kappa^\Theta_{i_h,+} = \dfrac{k\Delta n}{n_0} \sum_{m \in \mathbb{Z}^{(25)} } I^{m,\Theta}_{i,+} + \dfrac{k(\Delta n-\epsilon^
\Theta)}{n_0}I^{v,\Theta}_{i,+}.\end{equation}

For the integral $I^{m,\Theta}_{i,+}$: When $m>i$,
\begin{equation}\label{Ithetami+1}
\begin{aligned}
I^{m,\Theta}_{i,+} =&
\dfrac{2 \pi B\cdot B^\Theta_+ a}{\Gamma^2-(\Gamma^\Theta_+)^2} K_0(\Gamma\vert i-m \vert D_h)K_0(\Gamma^\Theta_+ R_m)(\Gamma I_0(\Gamma^\Theta_+ a)I_1(\Gamma a)-\Gamma^\Theta_+ I_1(\Gamma^\Theta_+ a)I_0(\Gamma a))\\
&+\dfrac{4\pi B\cdot B^\Theta_+  a}{\Gamma^2-(\Gamma^\Theta_+)^2}  \sum\limits_{q=1}^{\infty} (-1)^{q} K_q(\Gamma \vert i-m\vert D_h) K_q (\Gamma^\Theta_+ R_m) \cos(q\alpha_m) (\Gamma I_q(\Gamma^\Theta_+ a)I_{q-1}(\Gamma a)-\Gamma^\Theta_+ I_{q-1}(\Gamma^\Theta_+ a)I_q(\Gamma a)).
 \end{aligned}\end{equation}

When $m<i$,
\begin{equation}\label{Ithetami+2}
\begin{aligned}
I^{m,\Theta}_{i,+} =&
\dfrac{2 \pi B\cdot B^\Theta_+ a}{\Gamma^2-(\Gamma^\Theta_+)^2} K_0(\Gamma\vert i-m \vert D_h)
K_0(\Gamma^\Theta_+ R_m)
(\Gamma I_0(\Gamma^\Theta_+ a)I_1(\Gamma a)-\Gamma^\Theta_+ I_1(\Gamma^\Theta_+ a)I_0(\Gamma a))\\
&+\dfrac{4\pi B\cdot B^\Theta_+  a}{\Gamma^2-(\Gamma^\Theta_+)^2}  \sum\limits_{q=1}^{\infty} K_q(\Gamma \vert i-m\vert D_h) K_q (\Gamma^\Theta_+ R_m) \cos(q\alpha_m) (\Gamma I_q(\Gamma^\Theta_+ a)I_{q-1}(\Gamma a)-\Gamma^\Theta_+ I_{q-1}(\Gamma^\Theta_+ a)I_q(\Gamma a)). \end{aligned}
\end{equation}

When $m=i$,
\begin{equation}\label{Ithetami+3}
I^{m,\Theta}_{i,+} =\dfrac{2\pi AB^\Theta_+ a}{\Lambda^2+(\Gamma^\Theta_+)^2}K_0(\Gamma^\Theta_+ R_i)\Big(\Lambda J_1(\Lambda a)\,I_0(\Gamma^\Theta_+ a)+\Gamma^\Theta_+ J_0(\Lambda a)\,I_1(\Gamma^\Theta_+ a)\Big).
\end{equation}

The integral $I^{v,\Theta}_{i,+}$ can be expressed as
\begin{equation}\label{Ithetavi+}
\begin{aligned}
I^{v,\Theta}_{i,+} &=
 \dfrac{2\pi B\cdot B^\Theta_+ a}{\Gamma^2-(\Gamma^\Theta_+)^2}   K_0(\Gamma R_i) K_0(\Gamma^\Theta_+ (2D_v))
(\Gamma I_0(\Gamma^\Theta_+ a)I_1(\Gamma a)-\Gamma^\Theta_+ I_1(\Gamma^\Theta_+ a)I_0(\Gamma a))\\
&+\dfrac{4\pi B\cdot B^\Theta_+  a}{\Gamma^2-(\Gamma^\Theta_+)^2} \sum_{q=1}^{\infty}  K_q(\Gamma R_i) K_q(\Gamma^\Theta_+ (2D_v))
\cos\left(q \left(\alpha_i - \dfrac{\pi}{2}\right)\right)
(\Gamma I_q(\Gamma^\Theta_+ a)I_{q-1}(\Gamma a)-\Gamma^\Theta_+ I_{q-1}(\Gamma^\Theta_+ a)I_q(\Gamma a)).
 \end{aligned}
\end{equation}
\end{proposition}

\begin{proposition}\label{prop:68}
$\kappa^\Theta_{i_h,-}$ can be expressed as
\begin{equation}\label{kappathetai-}\kappa^\Theta_{i_h,-} = \dfrac{k\Delta n}{n_0} \sum_{m \in \mathbb{Z}^{(25)} } I^{m,\Theta}_{i,-} + \dfrac{k(\Delta n+\epsilon^
\Theta)}{n_0}I^{v,\Theta}_{i,-}.\end{equation}

For the integral $I^{m,\Theta}_{i,-}$: When $m>i$,
\begin{equation}\label{Ithetami-1}
\begin{aligned}
I^{m,\Theta}_{i,-} =&
\dfrac{2 \pi B\cdot B^\Theta_{-} a}{\Gamma^2-(\Gamma^\Theta_{-})^2} K_0(\Gamma\vert i-m \vert D_h)K_0(\Gamma^\Theta_{-} R_m)(\Gamma I_0(\Gamma^\Theta_{-} a)I_1(\Gamma a)-\Gamma^\Theta_{-} I_1(\Gamma^\Theta_{-} a)I_0(\Gamma a))\\
&+\dfrac{4\pi B\cdot B^\Theta_{-}  a}{\Gamma^2-(\Gamma^\Theta_{-})^2}  \sum\limits_{q=1}^{\infty} (-1)^{q} K_q(\Gamma \vert i-m\vert D_h) K_q (\Gamma^\Theta_{-} R_m) \cos(q\alpha_m) (\Gamma I_q(\Gamma^\Theta_{-} a)I_{q-1}(\Gamma a)-\Gamma^\Theta_{-} I_{q-1}(\Gamma^\Theta_{-} a)I_q(\Gamma a)).
\end{aligned}
\end{equation}

When $m<i$,
\begin{equation}\label{Ithetami-2}
\begin{aligned}
I^{m,\Theta}_{i,-} =&
\dfrac{2 \pi B\cdot B^\Theta_{-} a}{\Gamma^2-(\Gamma^\Theta_{-})^2} K_0(\Gamma\vert i-m \vert D_h)
K_0(\Gamma^\Theta_{-} R_m)
(\Gamma I_0(\Gamma^\Theta_{-} a)I_1(\Gamma a)-\Gamma^\Theta_{-} I_1(\Gamma^\Theta_{-} a)I_0(\Gamma a))\\
&+\dfrac{4\pi B\cdot B^\Theta_{-}  a}{\Gamma^2-(\Gamma^\Theta_{-})^2}  \sum\limits_{q=1}^{\infty} K_q(\Gamma \vert i-m\vert D_h) K_q (\Gamma^\Theta_{-} R_m) \cos(q\alpha_m) (\Gamma I_q(\Gamma^\Theta_{-} a)I_{q-1}(\Gamma a)-\Gamma^\Theta_{-} I_{q-1}(\Gamma^\Theta_{-} a)I_q(\Gamma a)). \end{aligned}\end{equation}

When $m=i$,
\begin{equation}\label{Ithetami-3}I^{m,\Theta}_{i,-} =\dfrac{2\pi AB^\Theta_{-} a}{\Lambda^2+(\Gamma^\Theta_{-})^2}K_0(\Gamma^\Theta_{-} R_i)\Big(\Lambda J_1(\Lambda a)\,I_0(\Gamma^\Theta_{-} a)+\Gamma^\Theta_{-} J_0(\Lambda a)\,I_1(\Gamma^\Theta_{-} a)\Big).
\end{equation}

The integral $I^{v,\Theta}_{i,-}$ can be expressed as:
\begin{equation}\label{Ithetavi-}
\begin{aligned}
I^{v,\Theta}_{i,-} &=
 \dfrac{2\pi B\cdot B^\Theta_{-}a }{\Gamma^2-(\Gamma^\Theta_{-})^2}   K_0(\Gamma R_i) K_0(\Gamma^\Theta_{-} (2D_v))
(\Gamma I_0(\Gamma^\Theta_{-} a)I_1(\Gamma a)-\Gamma^\Theta_{-} I_1(\Gamma^\Theta_{-} a)I_0(\Gamma a))\\
&+\dfrac{4\pi B\cdot B^\Theta_{-}  a}{\Gamma^2-(\Gamma^\Theta_{-})^2} \sum_{q=1}^{\infty}  K_q(\Gamma R_i) K_q(\Gamma^\Theta_{-} (2D_v))
\cos\left(q \left(\alpha_i - \dfrac{\pi}{2}\right)\right)
(\Gamma I_q(\Gamma^\Theta_{-} a)I_{q-1}(\Gamma a)-\Gamma^\Theta_{-} I_{q-1}(\Gamma^\Theta_{-} a)I_q(\Gamma a)),
 \end{aligned}
\end{equation}
\end{proposition}

\subsection{Numerical simulation}

\begin{proposition}
Quadratic quantity $(\bm{C}^t)^\dagger  \bm{S}^t_\Theta \bm {C}^t$ is preserved on $z$.
\end{proposition}
\begin{proof}By Proposition~\ref{SKThetasym}, the matrices $\bm{S}^t_\Theta$ and $\bm{K}^t_\Theta$ are Hermitian.
$$\ii\bm{S}^t_\Theta\dfrac{\mathrm{d}}{\mathrm{d}z}\bm{C}^t(z)+\bm{K}^t_\Theta\bm{C}^t(z)=\bm{0}\implies \bm{S}^t_\Theta\dfrac{\mathrm{d}}{\mathrm{d}z}\bm{C}^t(z)=\ii\bm{K}^t_\Theta\bm{C}^t(z)\text{ and }  \dfrac{\mathrm{d}}{\mathrm{d}z}(\bm{C}^t(z))^\dagger \bm{S}^t_\Theta=-\ii (\bm{C}^t)^\dagger \bm{K}^t_\Theta.$$
Then
$$\dfrac{\mathrm{d}}{\mathrm{d}z}((\bm{C}^t(z))^\dagger  \bm{S}^t_\Theta \bm {C}^t(z))=(\dfrac{\mathrm{d}}{\mathrm{d}z}(\bm{C}^t(z))^\dagger ) \bm{S}^t_\Theta \bm {C}^t(z)+   (\bm {C}^t(z))^\dagger\bm{S}^t_\Theta(\dfrac{\mathrm{d}}{\mathrm{d}z}\bm{C}^t(z))=-\ii (\bm{C}^t)^\dagger \bm{K}^t_\Theta \bm{C}^t+\ii (\bm{C}^t)^\dagger \bm{K}^t_\Theta\bm{C}^t=0$$
\end{proof}
\begin{definition}The {\bf total power} at a propagation distance $z$ for the mode envelope \eqref{psitheta} is defined as $P(z) := (\bm{C}^t(z))^\dagger \bm{S}^t_\Theta \bm{C}^t(z)$.
\end{definition}
Meanwhile, to distinguish the contributions from the horizontal waveguide array and the two vertically displaced waveguides, we define the sub-vectors $\bm{C}_H$ and $\bm{C}_V$ containing the corresponding amplitudes.
The horizontal power $P_H$ and vertical power $P_V$ are expressed as:
$$P_H(z) = \text{Re} \left[ \bm{C}_H^\dagger \bm{S}_{HH} \bm{C}_H + \bm{C}_H^\dagger \bm{S}_{HV} \bm{C}_V \right]$$
$$P_V(z) = \text{Re} \left[ \bm{C}_V^\dagger \bm{S}_{VV} \bm{C}_V + \bm{C}_V^\dagger \bm{S}_{VH} \bm{C}_H \right],$$
 where $\bm{S}_{HH}$ and $\bm{S}_{VV}$ are the diagonal blocks of the overlap matrix $\bm{S}^t_\Theta$ corresponding to the horizontal and vertical subsystems, respectively.
 $\bm{S}_{HV}$ represents the cross-overlap (coupling) matrix between horizontal and vertical modes, with $\bm{S}_{VH} = \bm{S}_{HV}^\dagger$.

Our approach utilizes the closed-form analytical expressions for the coupling and overlap elements. Specifically, the matrices $\bm{S}^t_\Theta$ and $\bm{K}^t_\Theta$ are constructed using the comprehensive set of overlap integrals:
\begin{itemize}
    \item \textbf{Intra-array interaction:} $I_{m,i,j}$ and $I^v_{ij}$ defining the continuum band.
    \item \textbf{Perturbed vertical coupling:} $I^{m,\Theta}_{\pm,\pm}$ and $I^{v,\Theta}_{\pm,\pm}$, which incorporate the symmetry-breaking parameter $\epsilon^\Theta$.
    \item \textbf{Vertical-Horizontal interaction:} The specific coupling terms $I^{m,\Theta}_{i,\pm}$ and $I^{v,\Theta}_{i,\pm}$ between the horizontal array and the vertical waveguides.
\end{itemize}
From Proposition \ref{prop:63}-\ref{prop:68}, we could notice that there is an infinite sum in each integral. In the numerical calculation, we calculate each integral with a truncation order of $Q_{max}=5$, i.e. calculate the sum till $q=5$. This ensures that the non-orthogonality of the leaky modes is captured with high fidelity without the accumulation of spatial integration errors.

\begin{algorithm}[H]
\caption{Detailed Numerical Implementation of BIC Leakage}
\begin{algorithmic}[1]
\Require $\epsilon^\Theta$, $\{a, D_h, D_v\}$, and Z-grid.
\Ensure Intensity evolution $|\psi^t_\Theta(x,y,z)|^2$.

\State \textbf{Step 1: Modal Analysis}
\State \quad Compute $\beta_0, \beta_0^{\Theta,\pm}$ and normalization constants $A^\Theta_{\pm}, B^\Theta_\pm$.

\State \textbf{Step 2: Semi-Analytical Integral Computation}
\State \quad \textbf{Horizontal-Horizontal:} Compute $I_{m,i,j}$ and $I^v_{ij}$ for waveguide array.
\State \quad \textbf{Self-Vertical:} Compute $I^{m,\Theta}_{+,+}, I^{v,\Theta}_{+,+}, I^{m,\Theta}_{-,-}, I^{v,\Theta}_{-,-}$ for the perturbed waveguides.
\State \quad \textbf{Cross-Vertical:} Compute $I^{m,\Theta}_{-,+}, I^{v,\Theta}_{-,+}$ to account for $\pm$ interaction.
\State \quad \textbf{V-H Coupling:} Compute $I^{m,\Theta}_{i,\pm}, I^{v,\Theta}_{i,\pm}$ for each $i \in \mathbb{Z}^{(25)}$.
\State \quad \textit{(Note: All integrals are evaluated via closed-form Bessel addition theorems with $Q_{max}=5$.)}

\State \textbf{Step 3: Matrix Assembly}
\State \quad Calculate $\bm{S}^t_\Theta$, $\bm{\kappa}^t_\Theta$ and $\bm{K}^t_\Theta$.
\State \textbf{Step 4: Unitary Propagation (The "Solver" Core)}
\State \quad Initialize $\bm{C}^t(0) = [0, \dots, 1, 0, -1, \dots, 0]^\mathsf{T}$.
\State \quad \textbf{Solve} the implicit system: $\left(\bm{S}^t_\Theta - \frac{\ii\Delta z}{2}\bm{K}^t_\Theta\right) \bm{C}^{n+1} = \left(\bm{S}^t_\Theta + \frac{\ii\Delta z}{2}\bm{K}^t_\Theta\right) \bm{C}^n$.

\State \textbf{Step 5: Field Reconstruction}
\State \quad Sum the modes: $\psi^t_\Theta(x,y,z) = \sum c_j(z) \phi_j(x,y)$.
\State \quad Compute intensity distribution $I(x,y,z) = |\psi(x,y,z)|^2$.
\State \Return $I(x,y,z)$ for visualization (Replicating Fig. 3b).
\end{algorithmic}
\end{algorithm}

We set the perturbation of the refractive index as $\epsilon^\Theta=8\times 10^{-5}$ in the numerical simulation. To numerically solve the coupled-mode system, we discretize the propagation coordinate $z$ using a Crank--Nicolson scheme\cite{crank1947practical}. The discretized evolution is given by
\begin{equation}
\bm{S}^t_\Theta\left(\bm{C}^{n+1} - \bm{C}^{n}\right)
=
\ii\,\frac{\Delta z}{2}\,
\bm{K}^t_\Theta\left(\bm{C}^{n+1} + \bm{C}^{n}\right).
\end{equation}

This implicit scheme is unconditionally stable and conserves quadratic invariants $ (\bm{C}^t)^\dagger  \bm{S}^t_\Theta \bm {C}^t$(see Hairer~\cite[p.~97, Theorem~2.1]{hairer2006structure}).

The initial condition is chosen as a normalized antisymmetric excitation,
$$
\bm{C}(0) = \frac{1}{\sqrt{2}}\bigl(0,\ldots,0,1,0,-1,0,\ldots,0\bigr)^{\mathsf{T}},
$$
which is selected to excite the symmetry-protected state.
\begin{figure}[htbp]
   \centering
   \includegraphics[width=\textwidth]{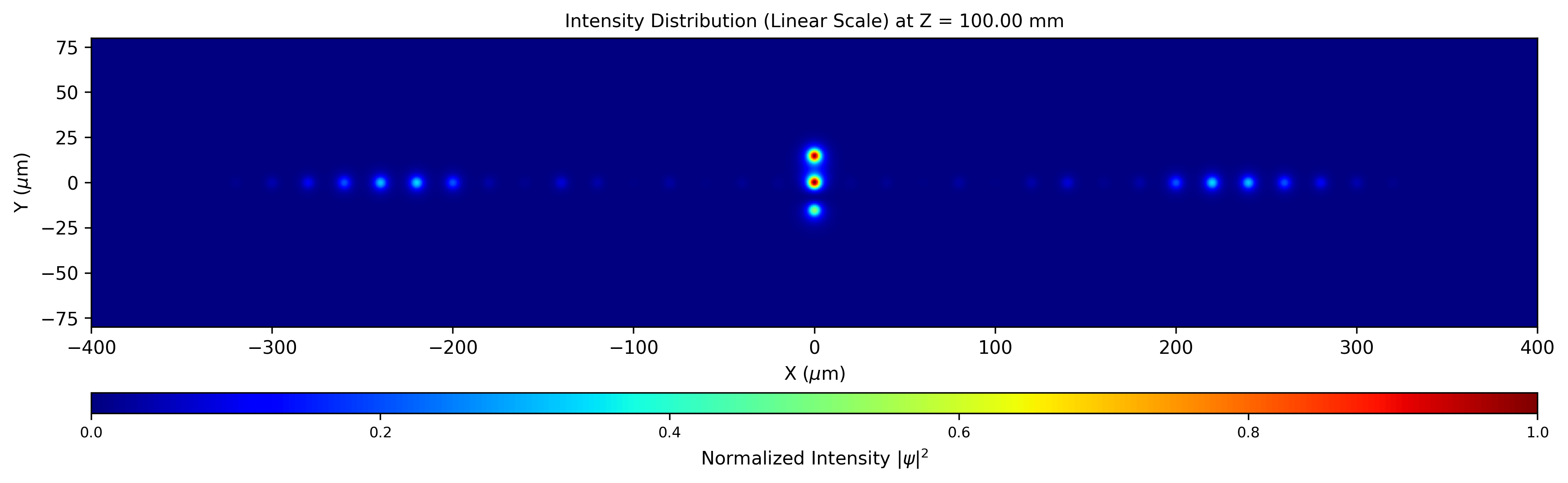} 
   \caption{Intensity Distribution at $z=100.00mm$}
   \label{Intensity Distribution at $z=100.00mm$}
\end{figure}
Numerical simulation results are presented at $z = 100$~mm, which is also the propagation length used in the experiment by Plotnik et al. \cite[p.~3]{Plotnik2011OpticalBIC}. The result confirms the significant leakage of the bound state due to symmetry breaking ($\epsilon^\Theta = 8 \times 10^{-5}$).

The diagnostics of mode amplitudes reveal a significant redistribution of power: the total power in the vertical waveguides, $P_V$, drops to $27.53\%$, while the horizontal continuum states capture the remaining $72.47\%$. These data prove that the perturbation successfully couples the discrete BIC to the lattice continuum, transforming the bound state into a leaky mode with a well-defined radiation pattern.

\section{Conclusion}
In this work, we have developed an analytical framework for the study of symmetry-protected Bound States in the Continuum (BICs) in both finite and infinite optical waveguide arrays. By adopting Nonorthogonal Coupled-Mode Equations (NCME) rather than the simplified tight-binding approximation, the model explicitly accounts for the non-vanishing modal overlaps between adjacent waveguides. Through a systematic application of addition theorems for Bessel functions, we derived exact analytical expressions for the entries of the overlap and coupling matrices. This semi-analytical formulation avoids the truncation errors inherent in numerical integration over unbounded domains and provides a high-precision foundation for capturing the delicate interference mechanisms underlying BIC formation.

By characterizing the infinite waveguide array through the Fourier symbol of the associated convolution operator on $\ell^2(\mathbb{Z})$ and invoking Wiener’s lemma within the framework of abstract harmonic analysis on Banach algebras, we obtained a strict spectral characterization of the horizontal continuum, together with precise conditions under which a discrete mode remains embedded without radiation. The transition from an ideal symmetry-protected BIC to a dissipative leaky mode was quantitatively described via a symmetry-breaking perturbation. Numerical simulations, performed using a power-conserving Crank–Nicolson scheme, confirmed that the localized discrete state couples efficiently to lattice radiation modes once the vertical parity symmetry is broken. These results demonstrate that the proposed NCME-based model achieves a favorable balance between computational efficiency and physical fidelity, outperforming standard orthogonal approximations.

Future work will aim at extending the present semi-analytical framework to more intricate geometries and anisotropic media. From a mathematical standpoint, a particularly promising direction lies in the application of layer potential techniques and boundary integral equation methods to the analysis of BICs. Layer potential theory could provide a more generalized framework for characterizing BICs in metamaterial structures with complex subwavelength architectures, further enhancing the robustness of computational photonics design.

\section*{Data availability}
The replication package contains  all data used in this research.

\section*{Declaration of competing interest}
The authors declare that they have no known competing financial interests or personal relationships that could have appeared to influence the work reported in this paper. 

\section*{Acknowledgements}
The research of W. W. was supported by NSFC grant (12301539) and youth growth grant of Department of Science and Technology of Jilin Province (20240602113RC).







\bibliographystyle{plain}
\bibliography{BIC1223}

\begin{thebibliography}{10}

\bibitem{bottcher2000toeplitz}
Albrecht B{\"o}ttcher and Sergei~M Grudsky.
\newblock {\em Toeplitz matrices, asymptotic linear algebra, and functional
  analysis}, volume~18.
\newblock Springer Science \& Business Media, 2000.

\bibitem{charalambides2011near}
Marcos Charalambides and Michael Christ.
\newblock Near-extremizers of young's inequality for discrete groups.
\newblock {\em arXiv preprint arXiv:1112.3716}, 2011.

\bibitem{chipot2005handbook}
Michel Chipot and Pavol Quittner.
\newblock {\em Handbook of Differential Equations: Stationary Partial
  Differential Equations}, volume~2.
\newblock Elsevier, 2005.

\bibitem{crank1947practical}
John Crank and Phyllis Nicolson.
\newblock A practical method for numerical evaluation of solutions of partial
  differential equations of the heat-conduction type.
\newblock In {\em Mathematical proceedings of the Cambridge philosophical
  society}, volume~43, pages 50--67. Cambridge University Press, 1947.

\bibitem{davies2007linear}
E~Brian Davies.
\newblock {\em Linear operators and their spectra}, volume 106.
\newblock Cambridge University Press, 2007.

\bibitem{edgar1979difference}
GA~Edgar and JM~Rosenblatt.
\newblock Difference equations over locally compact abelian groups.
\newblock {\em Transactions of the American Mathematical Society},
  253:273--289, 1979.

\bibitem{folland2016course}
Gerald~B Folland.
\newblock {\em A course in abstract harmonic analysis}.
\newblock CRC press, 2016.

\bibitem{hairer2006structure}
Ernst Hairer, Christian Lubich, and Gerhard Wanner.
\newblock Structure-preserving algorithms for ordinary differential equations.
\newblock {\em Geometric numerical integration}, 31, 2006.

\bibitem{haus2003coupled}
H~Haus, W~Huang, S~Kawakami, and N~Whitaker.
\newblock Coupled-mode theory of optical waveguides.
\newblock {\em Journal of lightwave technology}, 5(1):16--23, 2003.

\bibitem{hislop2012fundamentals}
Peter~D Hislop.
\newblock Fundamentals of scattering theory and resonances in quantum
  mechanics.
\newblock {\em Cubo (Temuco)}, 14(3):01--39, 2012.

\bibitem{horn2012matrix}
Roger~A Horn and Charles~R Johnson.
\newblock {\em Matrix analysis}.
\newblock Cambridge university press, 2012.

\bibitem{hsu2016bound}
Chia~Wei Hsu, Bo~Zhen, A~Douglas Stone, John~D Joannopoulos, and Marin
  Solja{\v{c}}i{\'c}.
\newblock Bound states in the continuum.
\newblock {\em Nature Reviews Materials}, 1(9):1--13, 2016.

\bibitem{huang1994coupled}
Wei-Ping Huang.
\newblock Coupled-mode theory for optical waveguides: an overview.
\newblock {\em Journal of the Optical Society of America A}, 11(3):963--983,
  1994.

\bibitem{katznelson2004introduction}
Yitzhak Katznelson.
\newblock {\em An introduction to harmonic analysis}.
\newblock Cambridge University Press, 2004.

\bibitem{Lifante2003IntegratedPhotonicsFundamentals}
Gin\'{e}s Lifante.
\newblock {\em Integrated Photonics: Fundamentals}.
\newblock John Wiley \& Sons, Inc., Chichester, West Sussex, England, 2003.

\bibitem{neumann1929merkwurdige}
J~von Neumann and Eugene Wigner.
\newblock {\"U}ber merkw{\"u}rdige diskrete eigenwerte.
\newblock {\em Phys. Z}, 30(524):291--293, 1929.

\bibitem{Okamoto2006Fundamentals}
Katsunari Okamoto.
\newblock {\em Fundamentals of Optical Waveguides}.
\newblock Academic Press, San Diego, 2 edition, 2006.

\bibitem{pazy2012semigroups}
Amnon Pazy.
\newblock {\em Semigroups of linear operators and applications to partial
  differential equations}, volume~44.
\newblock Springer Science \& Business Media, 2012.

\bibitem{petravcek2022bound}
J~Petr{\'a}{\v{c}}ek and V~Kuzmiak.
\newblock Bound states in the continuum in waveguide arrays within a symmetry
  classification scheme.
\newblock {\em Optics Express}, 30(20):35712--35724, 2022.

\bibitem{pinske2022symmetry}
Julien Pinske and Stefan Scheel.
\newblock Symmetry-protected non-abelian geometric phases in optical waveguides
  with nonorthogonal modes.
\newblock {\em Physical Review A}, 105(1):013507, 2022.

\bibitem{Plotnik2011OpticalBIC}
Y.~Plotnik, O.~Peleg, F.~Dreisow, M.~Heinrich, S.~Nolte, A.~Szameit, and
  M.~Segev.
\newblock Experimental observation of optical bound states in the continuum.
\newblock {\em Physical Review Letters}, 107(18):183901, 2011.

\bibitem{reed1972methods}
Michael Reed and Barry Simon.
\newblock {\em Methods of modern mathematical physics, 2. Fourier Analysis,
  Self-Adjointness}.
\newblock New York, London: Academic Press, 1972.

\bibitem{suh2004temporal}
Wonjoo Suh, Zheng Wang, and Shanhui Fan.
\newblock Temporal coupled-mode theory and the presence of non-orthogonal modes
  in lossless multimode cavities.
\newblock {\em IEEE Journal of Quantum Electronics}, 40(10):1511--1518, 2004.

\bibitem{von1993merkwurdige}
John von Neumann and Eugene~P Wigner.
\newblock {\"U}ber merkw{\"u}rdige diskrete eigenwerte.
\newblock In {\em The Collected Works of Eugene Paul Wigner: Part A: The
  Scientific Papers}, pages 291--293. Springer, 1993.

\bibitem{xia2011supermodes}
Cen Xia, Neng Bai, Ibrahim Ozdur, Xiang Zhou, and Guifang Li.
\newblock Supermodes for optical transmission.
\newblock {\em Optics express}, 19(17):16653--16664, 2011.

\bibitem{xia2015supermodes}
Cen Xia, M~Amin Eftekhar, Rodrigo~Amezcua Correa, Jose~Enrique Antonio-Lopez,
  Axel Sch{\"u}lzgen, Demetrios Christodoulides, and Guifang Li.
\newblock Supermodes in coupled multi-core waveguide structures.
\newblock {\em IEEE Journal of Selected Topics in Quantum Electronics},
  22(2):196--207, 2015.

\bibitem{yang2024programmable}
Yang Yang, Robert~J Chapman, Ben Haylock, Francesco Lenzini, Yogesh~N Joglekar,
  Mirko Lobino, and Alberto Peruzzo.
\newblock Programmable high-dimensional hamiltonian in a photonic waveguide
  array.
\newblock {\em Nature Communications}, 15(1):50, 2024.

\bibitem{yang2017approximating}
Zhen-Hang Yang and Yu-Ming Chu.
\newblock On approximating the modified bessel function of the second kind.
\newblock {\em Journal of inequalities and applications}, 2017(1):41, 2017.

\end{thebibliography}


\begin{appendix}
\section{Properties of Bessel functions}
In this part we introduce some properties of Bessel functions, which will be of later use. Recall that

\begin{lemma}\label{Besselint}
The following integral identities hold:
\begin{align}
&\int_{x_1}^{x_2} J_0^2(\Lambda r)  r  \mathrm{d}r = \frac{r^2}{2}\left.\left[J_0^2(\Lambda r) + J_1^2(\Lambda r)\right]\right|_{x_1}^{x_2}, \\
&\int_{x_1}^{x_2}I_n^2(\Gamma r)r\mathrm{d}r=\frac{r^2}{2}\left.\left[I_n^2(\Gamma r)-I_{n-1}(\Gamma r)I_{n+1}(\Gamma r)\right]\right|_{x_1}^{x_2},\label{eq:bessel2}\\
&\int_{x_1}^{x_2} K_0^2(\Gamma r) r \mathrm{d}r = \frac{r^2}{2}\left.\left[ K_0^2(\Gamma r)- K_1^2(\Gamma r) \right]\right|_{x_1}^{x_2}, \\
&\int_{x_1}^{x_2} J_0(\Lambda r) I_0(\Gamma r) r \, dr = \frac{r}{\Lambda^2 + \Gamma^2} \left.\left[\Lambda I_0(\Gamma r) J_1(\Lambda r) + \Gamma J_0(\Lambda r) I_1(\Gamma r)\right]\right|_{x_1}^{x_2},\label{eq:bessel4} \\
&\int_{x_1}^{x_2} K_0(\Gamma r)I_0(\Gamma r)r\mathrm{d}r =\frac{r^2}{2}\left.\left[I_0(\Gamma r)K_0(\Gamma r)+I_1(\Gamma r)K_1(\Gamma r)\right]\right|_{x_1}^{x_2},\label{eq:bessel5} \\
&\int_{x_1}^{x_2} K_0(A r) I_0(B r) r \, dr=\left[ -\frac{r}{A^2 - B^2} \left( A I_0(B r) K_1(A r) + B I_1(B r) K_0(A r) \right) \right]_{x_1}^{x_2}, \\
&\int_{x_1}^{x_2} K_0(A r) K_0(B r) r \, dr=\left[ \frac{r}{A^2 - B^2} \left( -A K_0(B r) K_1(A r) + B K_0(A r) K_1(B r) \right) \right]_{x_1}^{x_2},\\
&\int_0^{x_2} K_0(A r) I_0(B r) r \, dr=-\frac{x_2}{A^2 - B^2} \left( A I_0(B {x_2}) K_1(A {x_2}) + B I_1(B {x_2}) K_0(A {x_2}) \right) + \frac{1}{A^2 - B^2},\label{int0x2K0I0}\\
&\int_{x_1}^\infty K_0(A r) K_0(B r) r \, dr=-\frac{x_1}{A^2 - B^2} \left( -A K_0(B {x_1}) K_1(A {x_1}) + B K_0(A {x_1}) K_1(B {x_1}) \right),\label{intx1inftyK0K0}\\
&\int_{x_1}^{x_2} r\, I_n(A r)\, I_n(B r)\,dr=\frac{r}{B^2-A^2}\Big[B\, I_n(A r)\,I_{n-1}(B r)-A\, I_{n-1}(A r)\, I_n(B r)\Big]^{x_2}_{x_1}
\qquad (A\neq B).
\end{align}
\end{lemma}

\begin{lemma}[Addition theorem] \label{Additiontheorem}
For $0<r_1 < r_2$,
\begin{equation}\label{eq-3.18}
K_0(\Gamma \sqrt{r_1^2 + r_2^2 - 2 r_1 r_2 \cos \theta}) = \sum_{m=-\infty}^{\infty} I_m(\Gamma r_1) K_m(\Gamma r_2) \cos(m \theta),
\end{equation}
and \begin{equation}\label{eq-3.19}K_0(\Gamma \sqrt{r_1^2 + r_2^2 + 2 r_1 r_2 \cos \theta}) = \sum_{m=-\infty}^{\infty}(-1)^m I_m(\Gamma r_1) K_m(\Gamma r_2) \cos(m \theta).\end{equation}
\end{lemma}

\begin{lemma}\label{lem-3.8}For any fixed $x>0$ and positive integer$m$, one can bound the modified Bessel function of the first kind as
$$I_m(x)\leqslant \dfrac{x^m}{2^m\;m!}e^{x}.$$
\end{lemma}
\begin{lemma}\label{lem-propK_0leq}
For any $x>0$, the modified Bessel function of the second kind of order zero satisfies
$$
K_0(x)\leqslant \sqrt{\frac{\pi}{2x}}\,e^{-x}.
$$
(See Yang~\cite{yang2017approximating}).
\end{lemma}
\begin{lemma}\label{lem-3.9}For any fixed $x>0$ and integer $m\ge 1$, the modified Bessel function of the second kind satisfies
$$K_m(x)\leqslant \dfrac{1}{2}\bm{\Gamma}(m)\left(\dfrac{2}{x}\right)^{m}=\dfrac{1}{2}(m-1)!\left(\dfrac{2}{x}\right)^{m}.$$
\end{lemma}
\begin{proof}The integral representation of $K_m(x)$ is
$$K_m(x)=\dfrac{\sqrt{\pi}(x/2)^m}{\Gamma(m+1/2)}\int_1^\infty e^{-xt}(t^2-1)^{m-1/2}\mathrm{d}t.$$
For $t>1$, there holds the inequallity $$(t^2-1)^{m-1/2}\leqslant (t^2)^{m-1/2}=t^{2m-1}.$$
Thus, $$K_m(x)\leqslant \dfrac{\sqrt{\pi}(x/2)^m}{\Gamma(m+1/2)}\int_1^\infty e^{-xt}t^{2m-1}\mathrm{d}t.$$
Apply substitution $u=xt$, one obtains
$$\begin{aligned}\int_1^\infty e^{-xt}t^{2m-1}\mathrm{d}t=x^{-2m}\int_x^{\infty}e^{-u}u^{2m-1}\mathrm{d}u\leqslant x^{-2m}\int_0^\infty e^{-u}u^{2m-1}\mathrm{d}u=x^{-2m}\Gamma(2m).\end{aligned}$$
Then, $$K_m(x)\leqslant \sqrt{\pi}\;2^{-m}x^{-m}\dfrac{\Gamma(2m)}{\Gamma(m+1/2)}.$$
From the Legendre duplication formula
$$
\Gamma(m)\Gamma(m+\dfrac{1}{2})=2^{1-2m}\sqrt{\pi}\;\Gamma(2m)
$$
we obtain the upper bound of $K_m(x)$
$$K_m(x)\leqslant \dfrac{1}{2}\Gamma(m)\left(\dfrac{2}{x}\right)^{m}=\dfrac{1}{2}(m-1)!\left(\dfrac{2}{x}\right)^{m}.$$
\end{proof}

\end{appendix}
\end{document}